\documentclass[fleqn,usenatbib]{mnras}

\usepackage{newtxtext,newtxmath}
\usepackage{booktabs}
\usepackage{amsmath}
\usepackage{longtable}
\usepackage{pdflscape}
\usepackage{array}
\usepackage{stfloats}
\usepackage{balance}

\usepackage[T1]{fontenc}

\DeclareRobustCommand{\VAN}[3]{#2}
\let\VANthebibliography\thebibliography
\def\thebibliography{\DeclareRobustCommand{\VAN}[3]{##3}\VANthebibliography}

\usepackage{graphicx}	
\usepackage{amsmath}	

\title[Dominant dissipation in patchy coronae]{A narrow Comptonization locus in Seyfert slab coronae: implications for dominant coronal dissipation and reduced feedback}

\author[H. Xu]{
Haichao Xu
\thanks{E-mail: haichao\_xu@zju.edu.cn}
\\
Institute for Astronomy, School of Physics, Zhejiang University, 866 Yuhangtang Road,
Hangzhou 310058, China
}

\date{Accepted XXX. Received YYY; in original form ZZZ}

\pubyear{\the\year{}}

\begin{document}
\label{firstpage}
\pagerange{\pageref{firstpage}--\pageref{lastpage}}
\maketitle

\begin{abstract}

The thermal state of active galactic nucleus (AGN) coronae is commonly described by the electron temperature $kT_{\rm e}$, the Thomson optical depth $\tau_{\rm T}$, and the geometry of the Comptonizing medium. We compile a literature sample of radio-quiet Seyfert galaxies with direct slab-geometry thermal Comptonization measurements of $kT_{\rm e}$ and $\tau_{\rm T}$. For bottom-illuminated slab coronae, we adopt a geometry-motivated effective Comptonization parameter, $y=4\theta\tau_{\rm T}\,{K_3(1/\theta)} / {K_2(1/\theta)}$, where $\theta=kT_{\rm e}/m_{\rm e}c^2$ and $K_n$ is the modified Bessel function of the second kind. We find that the compiled slab-corona measurements are distributed along a narrow anti-correlated ridge in the $kT_{\rm e}-\tau_{\rm T}$ plane. For the cleaned primary \texttt{compTT} sample, this ridge corresponds to $\langle y\rangle=0.770$ with a logarithmic dispersion of only 0.086 dex. We then compare the observed ridge with reduced-feedback slab-equilibrium calculations and find that the locus favours a corona-dominated local energy budget, $f\simeq1$, together with a feedback factor well below the full-covering value, $g\ll1$. We interpret the constant-$y$ locus as implications for dominant local coronal dissipation and reduced disc-corona radiative feedback, consistent with a patchy corona. Supplementary pair-balance calculations indicate that the low-temperature, high-optical-depth part of the sample is unlikely to be sustained by a purely thermal pair plasma alone, suggesting that the fitted optical depths are more plausibly dominated by electron--ion plasma.

\end{abstract}




\section{Introduction}

The hard X-ray continuum in active galactic nuclei (AGNs) is widely attributed to inverse Compton scattering of soft seed photons from the accretion disc by a hot corona \citep[e.g.,][]{1979ApJ...229..318G,1980A&A....86..121S,1994ApJ...436..599S}. In this picture, the coronal temperature, optical depth, and the geometry jointly determine the efficiency of Comptonization and thus shape the emergent high-energy spectrum. A central quantity in this context is the Compton $y$-parameter, which links the microscopic scattering process to the macroscopic spectral properties and energy dissipation of the source \citep[e.g.,][]{1979rpa..book.....R}. Determining whether AGN coronae occupy a characteristic region in this parameter space, and what sets that region, is therefore important for understanding the thermal regulation of accreting black-hole coronae.

While X-ray coronae are often modelled as spherical or lamppost-like sources \citep[e.g.,][]{2013MNRAS.430.1694D,2015ApJ...814...24L}, recent \textit{IXPE} \citep{2022JATIS...8b6002W} polarimetry increasingly favours the picture that coronae are spatially extended along the accretion-disc plane in bright AGNs and X-ray binaries (XRBs) \citep[e.g.,][]{2022Sci...378..650K,2023MNRAS.523.4468G}. Polarimetry alone, however, does not uniquely distinguish between a sandwich/slab corona overlying a thin disc \citep[e.g.,][hereafter HM91, HM93, PS96]{1991ApJ...380L..51H, 1993ApJ...413..507H, 1996ApJ...470..249P} and a truncated-disc geometry with an inner advection-dominated accretion flow (ADAF) \citep[e.g.,][]{1994ApJ...428L..13N,1997ApJ...489..865E,2007A&ARv..15....1D}. Nevertheless, relativistic reflection spectroscopy provides an independent probe of the radial extent of the cold, optically thick reflector, and observations of a number of black-hole XRBs in bright hard states indicate that the inner disc can remain close to the innermost stable circular orbit (ISCO), rather than being strongly truncated \citep[e.g.,][]{2006ApJ...653..525M,2010MNRAS.402..836R,2023ApJ...951..145L}. Taken together, the polarimetric evidence for disc-plane extended coronae, the reflection-based constraints on weak disc truncation, and the expectation that ADAFs are generally relevant only at very low accretion rates \citep[e.g.,][]{1995ApJ...452..710N, 1997ApJ...489..865E} indicate that luminous AGNs accreting at moderate rates through standard Shakura-Sunyaev discs (SSDs) \citep{1973A&A....24..337S} are more naturally interpreted within an extended slab-corona framework.

In the classical sandwich geometry, a hot Comptonizing corona overlies a cold accretion disc (e.g., HM91; HM93). The disc supplies the soft seed photons for Comptonization, while also reprocessing and reflecting the downward hard X-ray radiation. The physical geometry of such a corona need not be a homogeneous, full-covering atmosphere \citep[e.g.,][]{1979ApJ...229..318G, 1994ApJ...432L..95H}. A patchy-corona picture can reduce the radiative feedback between the disc and the active coronal regions relative to a homogeneous full-covering sandwich slab \citep[e.g.,][also see PS96]{1994ApJ...432L..95H}. In such a configuration, coronal heating, inverse Compton cooling, and disc radiative feedback are tightly coupled. If the corona is patchy or composed of finite active regions, only a fraction of the reprocessed disc radiation returns to the Comptonizing plasma. As a result, the observable combinations of electron temperature and optical depth should not be distributed arbitrarily, but instead should be organized by the radiative equilibrium of the slab disc-corona system. With modern broadband X-ray spectroscopy, such as \textit{Swift} and \textit{NuSTAR} \citep{2004ApJ...611.1005G, 2013ApJ...770..103H}, these two microscopic parameters, $kT_{\rm e}$ and $\tau_{\rm T}$, can now be constrained much more directly than before \citep[e.g.,][]{2015MNRAS.451.4375F,2018A&A...614A..37T}, making the $kT_{\rm e}-\tau_{\rm T}$ plane a natural parameter space to search for such an equilibrium locus.

In this paper, we compile a literature sample of AGN coronal temperatures and optical depths constrained from broadband spectral fits. By restricting the sample to moderate-accretion-rate sources, we re-examine their distribution in the $kT_{\rm e}-\tau_{\rm T}$ plane under the slab-corona framework. For soft seed photons, we adopt an effective Compton $y$-parameter appropriate for bottom-illuminated slab geometry. We find the measurements occupy a nearly constant $y$ track in the $kT_{\rm e}-\tau_{\rm T}$ plane. We then compare this empirical locus with reduced-feedback slab radiative-equilibrium boundaries parameterised by the coronal dissipation fraction and the feedback factor. We show that reproducing the observed locus requires a corona-dominated local energy budget and feedback well below the homogeneous full-covering limit. These results suggest that luminous AGN coronae occupy a stable Comptonization branch regulated by slab radiative balance and provide a new observational constraint to how accretion power is partitioned between the disc and the corona.

\section{Data and Sample Selection} 
\label{sec:data}

To investigate the thermal equilibrium states of AGN coronae in the $kT_{\rm e}-\tau_{\rm T}$ plane, we compile a literature sample of radio-quiet Seyfert galaxies with broadband X-ray constraints on the electron temperature $kT_{\rm e}$ and Thomson optical depth $\tau_{\rm T}$. We require direct slab-geometry thermal Comptonization fits that report best-fitting values and uncertainties for both $kT_{\rm e}$ and $\tau_{\rm T}$. Sources for which only phenomenological high-energy cutoffs are available are not included in the primary sample, because converting $E_{\rm cut}$ into $kT_{\rm e}$ and then calculating $\tau_{\rm T}$ is model-, geometry-, and optical-depth-dependent.

The primary sample is based on measurements obtained with the \texttt{compTT} model \citep[e.g.,][]{1980A&A....86..121S, 1994ApJ...434..570T, 1995ApJ...449..188H}, which has been widely used to jointly constrain $kT_{\rm e}$ and $\tau_{\rm T}$ from broadband AGN spectra. Its main parent dataset is \citet{2024A&A...690A.145S} (hereafter S24), who analysed 46 \textit{NuSTAR} observations of 20 AGNs and reported coronal temperatures and optical depths using \texttt{compTT} for both slab and spherical geometries. We adopt their slab-geometry results as the basis of the primary sample. To keep the sample focused on radio-quiet Seyfert coronae and to avoid possible jet contamination of the hard X-ray continuum, we exclude the radio-loud sources from the parent sample.

A key issue is the optical-depth convention. In the slab implementation of \texttt{compTT}, the reported optical-depth parameter follows the half-thickness convention of the Titarchuk slab formalism \citep[e.g.,][]{1994ApJ...434..570T, 1997A&A...323..259G}. In contrast, the slab optical depth used in our calculation and in the \texttt{compPS}-style plane-parallel convention corresponds to the full vertical Thomson depth across the slab \citep[e.g.,][]{1996ApJ...470..249P}. Representative comparisons over the hard-X-ray band relevant to \textit{NuSTAR} show that the high-energy continua predicted by \texttt{compTT} and \texttt{compPS} are approximately consistent once the \texttt{compTT} slab optical depth is converted to the full-depth convention \citep[also see][]{2018A&A...614A..37T}. We therefore convert the published slab-\texttt{compTT} optical depths according to
\begin{equation}
    \tau_{\rm T} = 2\tau_{\rm compTT}.
\end{equation}
Throughout this paper, all slab optical depths quoted below are therefore expressed in this full-vertical-depth convention and the values listed in Table \ref{tab:comptt_sample} have already been converted to this convention.

In addition to the S24 sample, we include two radio-quiet Seyfert galaxies with high-quality published slab-\texttt{compTT} measurements, Ark 120 and ESO 323-G77. We adopt optical classifications, redshifts, and Eddington ratios from the BASS DR2 catalogue whenever available \citep{2022ApJS..261....2K}. The \textit{Swift}/BAT 70-month catalogue is used as an additional cross-check and hard-X-ray source classification \citep{2013ApJS..207...19B}. The adopted \texttt{compTT} sample is listed in Table \ref{tab:comptt_sample}. Detailed source-by-source comments are given in Appendix \ref{sec:appendixA}.

\begin{table}
\centering
\caption{Radio-quiet Seyfert sample fitted with slab \texttt{compTT}.}
\label{tab:comptt_sample}
\renewcommand{\arraystretch}{1.25}
\begin{tabular}{llccc}
\hline
Source & Type & $kT_{\rm e}$ (keV) & $\tau_{\rm T}$ & $\lambda_{\rm Edd}$ \\
\hline
1H 0419-577 & Sy1 & $14^{+2}_{-1}$ & $5^{+0.4}_{-0.6}$ & 0.145 \\
Ark 564$^{\ddagger}$$^{\ast}$ & NLSy1 & $15 \pm 2$ & $2.8 \pm 0.2$ & 2.45 \\
ESO 103-G35 & Sy1.9 & $17^{+18}_{-3}$ & $4.4^{+0.6}_{-1.4}$ & 0.098 \\
ESO 362-G18 & Sy1.5 & $18^{+14}_{-4}$ & $5^{+0.8}_{-1.8}$ & 0.06 \\
ESO 383-G18$^{\ddagger}$ & Sy2 & $7.3^{+0.5}_{-0.4}$ & $9.4 \pm 0.6$ & 1.41 \\
GRS 1734-292 & Sy1 & $13^{+2}_{-1}$ & $5.8 \pm 0.4$ & 0.085 \\
HE 1143-1810 & Sy1 & $36^{+64}_{-19}$ & $2.4^{+2.2}_{-1.6}$ & 0.24 \\
IC 4329A & Sy1.2 & $44^{+20}_{-10}$ & $2.2 \pm 0.6$ & 0.17 \\
MCG-5-23-16$^{\S}$ & Sy1.9 & $41 \pm 11$ & $1.8 \pm 0.4$ & 0.032 \\
MCG+8-11-11$^{\ddagger}$ & Sy1 & $110^{+140}_{-40}$ & $0.6 \pm 0.4$ & 1.41 \\
Mrk 110 & Sy1 & $35^{+15}_{-10}$ & $2.4^{+1}_{-0.8}$ & 0.389 \\
Mrk 509 & Sy1.2 & $17^{+2}_{-1}$ & $4.4 \pm 0.2$ & 0.11 \\
NGC 3281$^{\dagger}$ & Sy2 & $11^{+4}_{-2}$ & $7.4^{+1.6}_{-1.8}$ & 0.0091 \\
NGC 5506$^{\ast}$ & Sy1.9 & $510^{+250}_{-150}$ & $0.04 \pm 0.02$ & 0.052 \\
NGC 5728$^{\dagger}$ & Sy1.9 & $13 \pm 1$ & $10^{+4}_{-2}$ & 0.0053 \\
NGC 6814 & Sy1 & $60^{+24}_{-20}$ & $1.6^{+1.4}_{-0.6}$ & 0.02 \\
SWIFT J2127.4+5654 & NLSy1 & $33^{+37}_{-15}$ & $2^{+1.6}_{-1.2}$ & 0.06 \\
UGC 6728 & Sy1 & $28^{+16}_{-18}$ & $4 \pm 2.4$ & 0.269 \\
Ark 120 & Sy1 & $150^{+160}_{-75}$ & $0.3^{+0.4}_{-0.1}$ & 0.066 \\
ESO 323-G77 & Sy1.2 & $38 \pm 2$ & $2.8 \pm 0.2$ & 0.269 \\
\hline
\end{tabular}
\begin{minipage}{0.98\linewidth}
\footnotesize
\emph{Notes.} Columns list the adopted optical type, coronal electron temperature, full vertical slab optical depth, and Eddington ratio. For slab-\texttt{compTT} measurements, the optical depths are reported in the full vertical slab convention adopted in this work. Superscripts denote: $^{\ast}$ denotes measurements beyond the model’s nominal validity range; $^{\dagger}$ denotes low-accretion-rate sources with $\lambda_{\rm Edd}<0.01$; $^{\ddagger}$ denotes high-accretion-rate sources with $\lambda_{\rm Edd}>1$; and $^{\S}$ denotes special source with source-specific caveats.
\end{minipage}
\renewcommand{\arraystretch}{1.0}
\end{table}

We also compile a small set of published slab-\texttt{compPS} measurements for comparison, including NGC 4151, IC 4329A, and MCG--5--23--16. These points are not used to define the primary statistical sample. Since $y$ depends directly on $\tau_{\rm T}$, combining \texttt{compPS} and \texttt{compTT} measurements in the same statistical sample would introduce additional model-systematic scatter. We therefore retain the \texttt{compPS} results only as supplementary model-systematics checks; they are listed separately in Table \ref{tab:compps_sample}. Source-by-source comments are provided in Appendix \ref{sec:appendixA}.

\begin{table}
\centering
\caption{Radio-quiet Seyfert sample fitted with slab \texttt{compPS}.}
\label{tab:compps_sample}
\renewcommand{\arraystretch}{1.25}
\begin{tabular}{llccc}
\hline
Source & Type & $kT_{\rm e}$ (keV) & $\tau_{\rm T}$ & $\lambda_{\rm Edd}$ \\
\hline
NGC 4151 & Sy1.5 & $62 \pm 7$ & $1.3 \pm 0.1$ & 0.019 \\
NGC 4151 & Sy1.5 & $200^{+7}_{-8}$ & $0.31 \pm 0.01$ & 0.01 \\
IC 4329A & Sy1.2 & $48 \pm 13$ & $1.5 \pm 0.39$ & 0.17 \\
MCG-5-23-16$^{\S}$$^{\ast}$ & Sy1.9 & $26 \pm 2$ & $2.2 \pm 0.1$ & 0.032 \\
MCG-5-23-16$^{\S}$ & Sy1.9 & $26 \pm 1$ & $1.25 \pm 0.05$ & 0.038 \\
\hline
\end{tabular}
\begin{minipage}{0.98\linewidth}
\footnotesize
\emph{Notes.} Columns are the same as in Table \ref{tab:comptt_sample}. The optical depths are given in the slab optical-depth convention of the corresponding \texttt{compPS} analyses. Superscripts have the same meanings as in Table \ref{tab:comptt_sample}.
\end{minipage}
\renewcommand{\arraystretch}{1.0}
\end{table}

We do not include fitting results obtained with other complex models, such as \texttt{MoCA} \citep{2018A&A...619A.105T}. Although \texttt{MoCA} provides a useful Monte Carlo treatment of Comptonization and polarization, \texttt{MoCA}-based slab fits can yield optical depths that are not directly interchangeable with those obtained from \texttt{compTT}/\texttt{compPS}, and model comparisons show that the spectral differences can become important in some regions of parameter space. Moreover, in the case of Ark 120, \citet{2019A&A...623A..12M} found no statistically significant preference among phenomenological, \texttt{nthcomp}, \texttt{compTT}, and \texttt{MoCA} descriptions based on spectroscopy alone. Since the aim of this work is to test the distribution of AGN coronae in a consistently defined $kT_{\rm e}-\tau_{\rm T}$ plane, we exclude \texttt{MoCA}, \texttt{nthcomp}-derived, and cutoff-only estimates from the primary analysis.

To assess the applicability of the standard thin-disc sandwich-corona picture, we restrict the primary sample to the moderate-accretion regime $0.01\leq\lambda_{\rm Edd}\leq 1$. Low-accretion-rate sources with $\lambda_{\rm Edd}<0.01$ may host truncated discs or ADAF-like inner flows \citep[e.g.,][]{1997ApJ...489..865E,2022ApJ...932...97C}, while highly super-Eddington sources with $\lambda_{\rm Edd} > 1$ may deviate from the geometrically thin disc approximation, with geometric thickening and radial advection becoming important \citep[e.g.,][]{1988ApJ...332..646A}. These sources are retained in the full table for completeness but are flagged and excluded from the fiducial cleaned sample when deriving the main $y$-distribution and correlation statistics. After applying these cuts and excluding flagged model-validity cases, the fiducial cleaned \texttt{compTT} sample contains 14 sources. Sources with additional source-specific caveats are also flagged in Table \ref{tab:comptt_sample} and discussed individually in Appendix \ref{sec:appendixA}.

It is worth noting that \citet{2018MNRAS.480.1819R} (hereafter R18) investigated the statistical properties of coronal parameters based on a large \textit{Swift}/BAT \citep{2005SSRv..120..143B} AGN sample. Their optical depth is not a direct spectral-fit parameter, but is inferred from a calibration between $\Gamma$, $kT_{\rm e}$ and $\tau_{\rm T}$ based on slab-\texttt{compPS} simulations. This calibration assumes $kT_{\rm e}=E_{\rm C}/2$ and a power-law approximation to the simulated 2--10 keV spectrum. Although this approach is useful for large-sample trends, it yields an effective ensemble optical depth rather than homogeneous, source-by-source Comptonization measurements. We therefore treat the R18 median point ($kT_{\rm e} = 105 \pm 18 \text{ keV}$ and $\tau_{\rm T} = 0.25 \pm 0.06$) only as an independent external population-level reference and do not include it in the statistics.

\section{Effective Compton \texorpdfstring{$y$}{y}-parameter for slab coronae}
\label{sec:slab_y}

In thermal Comptonization theory, the Compton $y$-parameter is commonly defined as the product of the mean fractional energy gain per scattering and the mean number of scatterings a photon undergoes before escape \citep[e.g.,][]{1979rpa..book.....R}:
\begin{equation}
y \simeq \left\langle \frac{\Delta \epsilon}{\epsilon} \right\rangle \langle N_{\rm sca} \rangle,
\label{eq:y-definition}
\end{equation}
where $\epsilon = E / m_{\rm e}c^2$ is the dimensionless photon energy.

For classical radiative transfer problems involving isotropic volume emission within a uniform plasma cloud, $\langle N_{\rm sca} \rangle$ scales approximately as $\tau_{\rm T}$ in the optically thin limit, while a random-walk approximation yields a $\tau_{\rm T}^2$ scaling in the optically thick regime \citep[e.g.,][]{1962ApJ...135..195O}. Consequently, approximate prescriptions such as $\tau_{\rm T}$, $\max (\tau_{\rm T},\tau_{\rm T}^2)$, or $\tau_{\rm T}+\tau_{\rm T}^2$ are often adopted in the literature. Meanwhile, the mean fractional energy gain $\langle\Delta \epsilon/\epsilon\rangle$ approaches $4\theta$ and $16\theta^2$ in the non-relativistic and ultra-relativistic limits, respectively. It is therefore commonly approximated as either $4\theta$ or the interpolation form $4\theta+16\theta^2$, where $\theta = kT_{\rm e} / m_{\rm e}c^2$ is the dimensionless electron temperature.

However, in the standard sandwich disc-corona paradigm, seed photons are not generated isotropically within the corona. Instead, they are injected upwards from the surface of the underlying cold accretion disc. This anisotropic, bottom-illuminated boundary condition differs fundamentally from the spherically symmetric diffusion problem associated with classical volume sources. Therefore, the effective Compton $y$-parameter for slab coronae should be defined with this boundary condition in mind.

\citet{2026arXiv260601087X} studied photon transport in slab geometry under the Thomson-scattering approximation. For Lambert-law injection from the lower boundary, appropriate for a locally thermal disc surface, the mean scattering number satisfies the exact relation
\begin{equation}
\langle N_{\rm sca}\rangle_{\rm L}=2\tau_{\rm T},
\label{eq:nsca_lambert}
\end{equation}
for arbitrary optical depth. Here, $\tau_{\rm T}$ denotes the full vertical optical depth of the slab. For normally incident beam injection, the mean scattering number also remains linear in the optically thin and optically thick limits, with the thick-limit behaviour $\langle N_{\rm sca}\rangle_{\rm b}\simeq 2.5\tau_{\rm T}$. Thus, for bottom-illuminated slab coronae, the main geometrical modification is the linear dependence on the full vertical optical depth, rather than the diffusion-like $\tau_{\rm T}^2$ dependence.

We next specify the energy-gain factor. In the soft-photon limit, the mean fractional energy gain in a scattering event by a single electron of Lorentz factor $\gamma$ is \citep[e.g.,][]{1970RvMP...42..237B, 1983ASPRv...2..189P}
\begin{equation}
\left\langle{\frac{\Delta\epsilon}{\epsilon}}\right\rangle_\gamma
\simeq\frac{4}{3}\gamma^2\beta^2 .
\end{equation}
Averaging this expression over a Maxwell-Jüttner electron distribution gives
\begin{equation}
     \left\langle{\frac{\Delta\epsilon}{\epsilon}}\right\rangle = 4\theta\frac{K_3(1/\theta)}{K_2(1/\theta)} = 4\theta \frac{K_1(1/\theta)}{K_2(1/\theta)} + 16\theta^2,
     \label{eq:enhancement_theo}
\end{equation}
where $\theta=kT_{\rm e}/m_{\rm e}c^2$, and $K_n$ is the modified Bessel function of the second kind. This expression reduces to the familiar non-relativistic result $4\theta$ at low temperature and includes the leading mildly relativistic correction. The simpler approximation $4\theta + 16\theta^2$ is commonly adopted in the literature and remains reasonably accurate in the asymptotic limits. Over the full temperature range, its deviation from the Maxwell-Jüttner averaged expression remains below $20\%$.

Motivated by the linear slab transport scaling, but retaining the conventional normalization of the classical $4\theta\tau_{\rm T}$ form, we define
\begin{equation}
y_{\rm slab} = 4\theta\frac{K_3(1/\theta)}{K_2(1/\theta)}
\tau_{\rm T}.
\label{eq:y_slab}
\end{equation}
Relative to the non-relativistic $4\theta\tau_{\rm T}$ form, this definition replaces the energy-gain factor with the Maxwell-Jüttner averaged one. If (\ref{eq:y-definition}) were applied literally with the Lambert-law scattering count, the resulting value would be larger by a constant factor of 2. A normally incident beam would change this normalization by a comparable factor, without changing the linear dependence on $\tau_{\rm T}$. Thus, $y$ should be regarded as a classically normalized effective diagnostic for comparing slab-corona measurements, rather than as a literal photon-counting estimate of the total fractional energy exchange. This constant normalization does not affect the source-to-source comparison, but should be kept in mind when comparing the absolute value of $y$ with calculations based directly on (\ref{eq:y-definition}). For comparison, we also evaluate three commonly used expressions: $4\theta \tau_{\rm T}$, $4\theta(1+4\theta)\tau_{\rm T}$, and $4\theta(1+4\theta)\tau_{\rm T}(1+\tau_{\rm T})$.

\section{Radiative equilibrium boundary in slab coronae}
\label{sec:boundary}

In the classical HM91/HM93 sandwich disc-corona picture, the gravitational energy released by accretion is dissipated partly in the cold disc and partly in the hot corona. The disc and the corona constitute a coupled two-phase system through the injection of soft seed photons, inverse Compton scattering, and the reflection and reprocessing of downward hard radiation.

Let $Q_{\rm diss}$ be the dissipated gravitational power, of which a fraction $f$ is transported to the corona to heat the plasma, while the remaining fraction $1-f$ is released internally within the disc. The coronal heating power is therefore
\begin{equation}
L_{\rm c}=fQ_{\rm diss}.
\end{equation}
We denote the soft photon luminosity entering the corona as $L_{\rm s}$. In a uniform full-covering sandwich slab, all disc photons enter the corona. More generally, following PS96, we introduce an effective feedback factor $g$, which parameterises the suppression of the radiative feedback from the irradiated disc to the active coronal region. The full-covering slab limit corresponds to $g=1$, while $g<1$ describes reduced radiative feedback, as expected for patchy coronal active regions.

A fraction of the Comptonized luminosity escapes downward and irradiates the accretion disc. We define the downward fraction of the Comptonized emission as
\begin{equation}
     \eta = \frac{L^{\downarrow}_{\rm comp}}{L^{\downarrow}_{\rm comp}+L^{\uparrow}_{\rm comp}}, 
\end{equation}
where $L^{\downarrow}_{\rm comp}$ and $L^{\uparrow}_{\rm comp}$ are the downward and upward escaping Comptonized luminosities, respectively. The irradiated disc responds through reflection and reprocessing. We define the disc albedo as
\begin{equation}
     a = \frac{L_{\rm refl}}{L_{\rm refl}+L_{\rm repr}},
\end{equation}
where $L_{\rm refl}$ and $L_{\rm repr}$ are the reflected and reprocessed luminosities, satisfying
\begin{equation}
     L_{\rm refl} + L_{\rm repr} = L_{\rm comp}^{\downarrow}.
\end{equation}
Only a fraction $g$ of these reflected and reprocessed photons is assumed to re-enter the active coronal region.

In a more rigorous radiative transfer treatment, we distinguish between the direct coronal heating power $L_{\rm c}$ and the total Comptonized hard luminosity $L_{\rm h}$, since the latter includes not only the energy dissipated in the corona but also the energy carried by scattered soft photons (e.g., PS96). We define $L_{\rm h}$ as the total Comptonized hard luminosity, i.e., the sum of the upward and downward scattered components ${L^{\downarrow}_{\rm comp}+L^{\uparrow}_{\rm comp}}$, and $p_{\rm sc}$ as the fraction of upward-injected soft photons that are actually scattered in the corona. The two-phase energy balance can then be written as
\begin{equation}
     \begin{aligned}
          L_{\rm s} &= g(1 - a)\eta L_{\rm h} + (1-f) Q_{\rm diss}, \\
          L_{\rm h} &= fQ_{\rm diss} + p_{\rm sc}[L_{\rm s} + ga\eta L_{\rm h}].
     \end{aligned}
\end{equation}

The first equation indicates that the soft luminosity entering the corona consists of locally produced disc photons plus the thermalized reprocessed radiation that returns to the active coronal region. The second equation indicates that the total Comptonized luminosity includes both direct coronal heating and the energy carried by soft photons that are scattered in the corona, including the returning reflected component. Solving these equations gives
\begin{equation}
    \begin{aligned}
\frac{L_{\rm s}}{Q_{\rm diss}} &=
\frac{g(1-a)\eta f + (1-f)(1-ga\eta p_{\rm sc})}
{1-p_{\rm sc}g\eta}, \\
\frac{L_{\rm h}}{Q_{\rm diss}} &=
\frac{f+p_{\rm sc}(1-f)}
{1-p_{\rm sc}g\eta}.
\end{aligned}
\end{equation}

Following PS96, we introduce the parameter $d$ to represent the ratio of the power dissipated in the disc to that in the corona, i.e.,
\begin{equation}
     d = \frac{1-f}{f}.
\end{equation}
The Compton amplification factor $A$, defined as the ratio of the hard luminosity to the soft luminosity, is given by:
\begin{equation}
     A = \frac{L_{\rm h}}{L_{\rm s}} = \frac{1 + p_{\rm sc}d}{g(1-a)\eta + d(1 - ga\eta p_{\rm sc})}.
     \label{eq:equi_eq}
\end{equation}
For given disc-corona energy partition fraction $f$ (or equivalently $d$) and feedback factor $g$, this expression uniquely specifies the Compton amplification required by the two-phase energy balance once $a$, $\eta$ and $p_{\rm sc}$ are fixed by the radiative transfer solution.

To evaluate the radiative-transfer quantities entering the energy-balance equation, we use the \texttt{compPSc} code \citep{2024ApJ...962..101Z}, a convolution version of the \texttt{compPS} iterative-scattering Comptonization model, to solve the anisotropic Compton radiative transfer in slab geometry for given $kT_{\rm e}$ and $\tau_{\rm T}$. We increase the maximum scattering order in \texttt{compPSc} to improve the convergence of the Comptonized spectrum in the mildly optically thick regime. From the resulting transfer solution, we extract the calculated quantities $A^{\rm cal}$, $\eta^{\rm cal}$, and $p^{\rm cal}_{\rm sc}$. 

For the disc response to downward hard radiation, we compute the effective albedo $a^{\rm cal}$ using the convolution model \texttt{ireflect} \citep{1995MNRAS.273..837M}. For each $kT_{\rm e}$ and $\tau_{\rm T}$, the downward Comptonized spectrum $L_{\rm comp}^\downarrow$ obtained from \texttt{compPSc} is used as the incident spectrum on the disc. The resulting reflected continuum is then integrated to estimate the reflected luminosity, while the difference between the incident and reflected luminosities gives the thermalized reprocessed component. Although \texttt{ireflect} is less detailed than modern reflection models such as \texttt{xillver} \citep{2010ApJ...718..695G, 2013ApJ...768..146G} or \texttt{relxill} \citep{2014ApJ...782...76G, 2014MNRAS.444L.100D}, it is sufficient for the present purpose because we require only the broadband reflected continuum and the corresponding energy-integrated albedo, rather than a detailed calculation of narrow Fe-line or relativistic reflection features. A practical advantage is that we can apply it directly to the full downward Comptonized spectrum computed by \texttt{compPSc}, rather than assuming a simple power-law, \texttt{nthcomp} or blackbody incident spectrum. The dependence of $A^{\rm cal}$, $\eta^{\rm cal}$, $p_{\rm sc}^{\rm cal}$, and $a^{\rm cal}$ on $kT_{\rm e}$ and $\tau_{\rm T}$, together with the numerical convergence region of the calculation, is shown in Appendix \ref{sec:appendixB}.

For any given $f$ and $g$, the two-phase energy balance defines the amplification required for radiative equilibrium (\ref{eq:equi_eq}), i.e. $A^{\rm req}(\eta^{\rm cal}, p^{\rm cal}_{\rm sc}, a^{\rm cal})$, while \texttt{compPSc} provides the actual radiative amplification $A^{\rm cal}$ produced at the corresponding $(\theta, \tau_{\rm T})$. By varying $\tau_{\rm T}$ at fixed $\theta$, we can numerically find the root where these two amplification factors coincide, thereby tracing the radiative equilibrium boundary $\tau_{\rm T}(\theta)$ for each choice of $f$ and $g$.

In our implementation, the Compton reflection hump produced by the cold disc is used only to modify the macroscopic disc-corona energy partition and is not fed back into \texttt{compPSc} as an additional bottom-boundary seed component for a second round of radiative transfer. We tested this approximation for several representative parameter sets and found that the additional feedback from repeatedly scattered reflection-hump photons produces only small shifts in the equilibrium boundaries, because the effective albedo remains relatively low over most of the parameter space.

\section{Results}

\begin{figure*}
\includegraphics{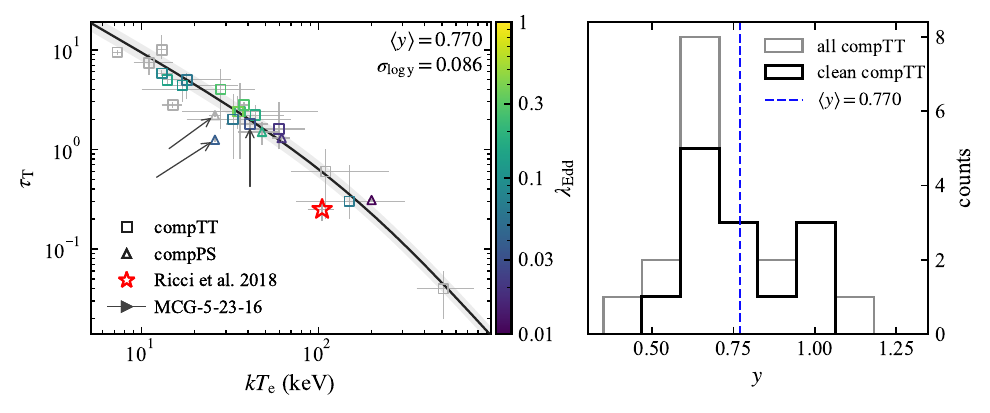}
\caption{Distribution of the compiled slab-corona measurements in the $kT_{\rm e}-\tau_{\rm T}$ plane and the corresponding $y$ distribution. Left: electron temperature versus full vertical Thomson optical depth. Optical depths of \texttt{compTT} measurements are converted to the full-thickness slab convention described in Sec. \ref{sec:data}. The \texttt{compPS} measurements are plotted for comparison but not included in the primary statistics. The red star marks the population-level median estimate from \citet{2018MNRAS.480.1819R}. Colours indicate the Eddington ratio $\lambda_{\rm Edd}$. Grey symbols denote sources retained in the full table but excluded from the fiducial cleaned sample because they are outside the adopted accretion-rate range $(0.01\leq\lambda_{\rm Edd}\leq1)$ or lie beyond the nominal model-validity range. The black curve shows the mean constant-$y$ locus of the cleaned \texttt{compTT} sample, and the shaded band indicates the corresponding logarithmic scatter. The arrows mark the measurements associated with special source MCG--5--23--16. Right: distribution of $y$ for the full and cleaned \texttt{compTT} samples.}
\label{fig:fitting}
\end{figure*}

Figure \ref{fig:fitting} shows the compiled slab-corona measurements in the $kT_{\rm e}-\tau_{\rm T}$ plane. The primary \texttt{compTT} sample shows a clear anti-correlation between electron temperature and optical depth: lower-temperature coronae have larger optical depths, while hotter coronae are optically thinner. This trend is not driven by the excluded objects. The grey points, which are retained in the full compilation but excluded from the fiducial cleaned sample because of accretion-regime or source-specific caveats, mostly follow the same broad locus, although they increase the total scatter. The supplementary \texttt{compPS} points are also broadly consistent with the same region of the $kT_{\rm e}-\tau_{\rm T}$ plane, but we do not include them in the primary statistics because of the additional model-systematic uncertainty. MCG--5--23--16 is a notable exception: the different published \texttt{compPS} measurements of this source span a range of optical depths and do not lie exactly on the cleaned \texttt{compTT} ridge.

The cleaned \texttt{compTT} sample ($N = 14$) is well described by a narrow constant-$y$ locus. Using the definition in (\ref{eq:y_slab}), we find
\begin{equation}
\langle y\rangle = 0.770,
\qquad
\sigma_{\log y}=0.086 ,
\end{equation}
where the scatter is measured as the standard deviation of $\log_{10} y$. The right panel of Fig. \ref{fig:fitting} shows that applying the cleaning criteria narrows the $y$ distribution but does not create the trend: the full \texttt{compTT} sample still remains concentrated around the same broad range of $y$, while the excluded objects mainly contribute to the tails of the distribution. The narrowness of the cleaned distribution indicates that radio-quiet Seyfert slab coronae do not occupy the $kT_{\rm e}-\tau_{\rm T}$ plane randomly, but instead lie close to a one-dimensional Comptonization locus.

We also tested how the concentration of the locus depends on the adopted definition of the Compton parameter. For the commonly used mildly relativistic approximation $4\theta(1+4\theta)\tau_{\rm T}$, we obtain $\langle y\rangle=0.829$ and $\sigma_{\log y}=0.091$. For the non-relativistic form $4\theta\tau_{\rm T}$, we obtain $\langle y\rangle=0.649$ and $\sigma_{\log y}=0.100$. Both definitions still reveal a narrow locus, because they retain the linear dependence on optical depth. In contrast, the expression $4\theta(1+4\theta)\tau_{\rm T}(1+\tau_{\rm T})$ gives a much broader distribution, with $\langle y\rangle=3.414$ and $\sigma_{\log y}=0.182$. This larger scatter is expected: for bottom-illuminated slab coronae, the mean scattering number scales approximately linearly with vertical optical depth, rather than as a diffusion-like $\tau_{\rm T}+\tau_{\rm T}^2$ prescription commonly associated with isotropic volume-source geometries \citep{2026arXiv260601087X}. The small difference between our fiducial $y_{\rm slab}$ and $4\theta(1+4\theta)\tau_{\rm T}$ reflects the fact that the interpolation $4\theta(1+4\theta)$ closely approximates the Maxwell-Jüttner averaged energy-gain factor $4\theta K_3(1/\theta)/K_2(1/\theta)$ over most of the temperature range relevant to the sample. Their largest relative difference occurs in the mildly relativistic regime around $\theta\sim1$, where it remains below $20\%$.

However, the population-level estimate from R18 lies below the cleaned \texttt{compTT} constant-$y$ ridge. We do not interpret this offset as a direct contradiction, because the R18 optical depth is not a source-by-source Comptonization fit parameter. It is an effective ensemble value inferred from a calibration between $\Gamma$, $kT_{\rm e}$, and $\tau_{\rm T}$ based on slab-\texttt{compPS} simulations, together with the assumption $kT_{\rm e}=E_{\rm C}/2$. The offset therefore illustrates the systematic uncertainty introduced when cutoff-based or calibration-derived optical depths are compared with direct \texttt{compTT}/\texttt{compPS} measurements. For this reason, the R18 point is shown only as an external population-level reference and is not included in the statistical analysis.

\begin{figure*}
\includegraphics{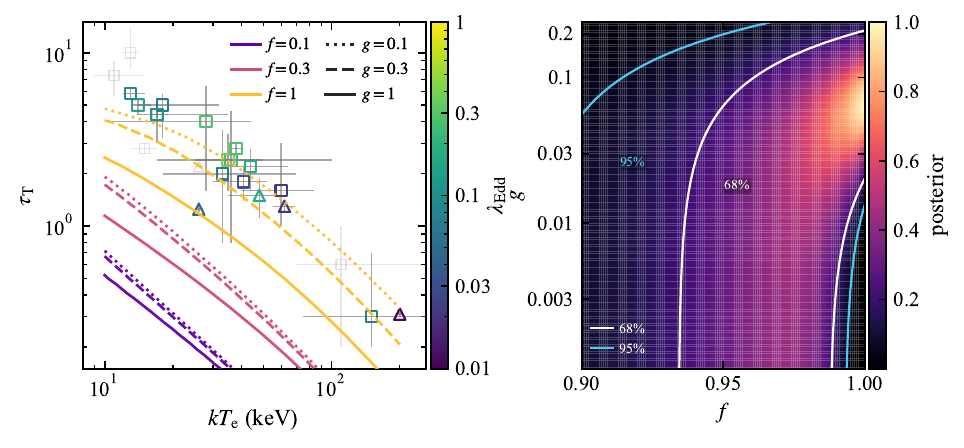}
\caption{Reduced-feedback slab-corona equilibrium and constraints on the energy-partition and feedback parameters ($f$ and $g$). Left: comparison between the observed $kT_{\rm e}-\tau_{\rm T}$ distribution and radiative-equilibrium boundaries computed within the reduced-feedback sandwich disc-corona framework. The symbols are the same as in Figure \ref{fig:fitting}. Colours of the curves indicate the local coronal dissipation fraction $f$, while line styles indicate the feedback factor $g$. Right: posterior distribution in the $f-g$ plane obtained by fitting the cleaned \texttt{compTT} sample with the sandwich disc-corona model. White and cyan contours mark the 68\% and 95\% credible regions, respectively.}
\label{fig:model}
\end{figure*}

Figure \ref{fig:model} compares the observed locus with the reduced-feedback slab-equilibrium boundaries described in Section \ref{sec:boundary}. The full-covering limit corresponds to $g=1$, whereas $g<1$ represents reduced radiative feedback between the cold disc and the active coronal region. The equilibrium curves show that decreasing $f$ moves the solutions away from the observed ridge: models in which only a modest fraction of the locally dissipated power is released in the corona systematically underpredict the optical depths of the observed sources at a given temperature. Increasing $f$ moves the equilibrium boundaries closer to the data, but a high coronal dissipation fraction alone is not sufficient if the feedback remains close to the homogeneous full-covering slab value.

The right panel of Figure \ref{fig:model} shows the posterior distribution in the $f-g$ plane obtained by fitting the cleaned \texttt{compTT} sample. The details of the likelihood construction and the adopted intrinsic-scatter treatment are given in Appendix \ref{sec:appendixC}. The posterior is elongated, reflecting the degeneracy between the local energy partition and the feedback strength. Nevertheless, the preferred region is confined to high coronal dissipation fractions, $f\simeq1$, together with feedback factors well below the full-covering value, $g=1$. Thus, within the reduced-feedback slab-equilibrium framework, the observed constant-$y$ locus favours a corona-dominated local energy budget and weakened disc-corona radiative feedback, which is naturally consistent with a patchy corona.

Finally, we tested whether the electron temperature and the effective Compton $y$ parameter of the sources are primarily driven by accretion rate. We performed Spearman rank correlation tests between $kT_{\rm e}$ and $\lambda_{\rm Edd}$, and between $y$ and $\lambda_{\rm Edd}$, using the central values of the \texttt{compTT} measurements. For the cleaned \texttt{compTT} sample, neither correlation is statistically significant: the coefficients are $\rho=-0.146$ with $p=0.620$ for $kT_{\rm e}-\lambda_{\rm Edd}$, and $\rho=0.359$ with $p=0.207$ for $y-\lambda_{\rm Edd}$. The same conclusion holds for the full \texttt{compTT} sample. In this case, we obtain $\rho=-0.027$ with $p=0.910$ for $kT_{\rm e}-\lambda_{\rm Edd}$, and $\rho=-0.156$ with $p=0.512$ for $y-\lambda_{\rm Edd}$. Although the cleaned sample shows a weak positive rank coefficient between $y$ and $\lambda_{\rm Edd}$, it is not statistically significant. The narrow $kT_{\rm e}-\tau_{\rm T}$ ridge is therefore more naturally interpreted as a slab-corona Comptonization locus regulated by radiative balance and disc-corona energy partition, rather than as a simple accretion-rate sequence.

\section{Discussion}

In this work, we find that the slab coronae of radio-quiet Seyfert galaxies do not populate the $kT_{\rm e}-\tau_{\rm T}$ plane randomly, but instead lie along a narrow anti-correlated locus. After converting the \texttt{compTT} slab optical depths to a common full-thickness convention, this locus is well described by a nearly constant effective slab Compton parameter, $y$. The small logarithmic dispersion of the cleaned sample indicates that the thermal state of the slab corona is not set solely by source-to-source stochastic variations, but is instead organized by a common Comptonization condition. In this sense, the observed ridge can be interpreted as a preferred slab-corona branch plausibly regulated by radiative balance.

A useful precedent for this interpretation is provided by the state-resolved Comptonization analysis of NGC 4151 \citep{2010MNRAS.408.1851L}. In that source, the fitted $kT_{\rm e}$ and $\tau_{\rm T}$ change substantially between bright and dim hard-X-ray states, and are strongly anti-correlated, whereas the Compton parameter remains approximately stable. This behaviour illustrates that $kT_{\rm e}$ and $\tau_{\rm T}$ are not always the most robust independent diagnostics of the coronal thermal state. Instead, the spectral curvature is often more directly related to a combined Comptonization measure $y$. The present multi-source sample can be viewed as an extension of this idea: a coherent ridge is observed across different AGN samples and different flux states. We interpret the persistence of the same narrow $y$ locus across heterogeneous sources and states as evidence that the ridge is not solely a parameter-degeneracy artefact, but reflects an underlying preferred Comptonization branch of slab-like AGN coronae. As discussed in Appendix \ref{sec:appendixC}, the inferred $f-g$ posterior remains qualitatively unchanged even when representative negative $kT_{\rm e}-\tau$ correlations are imposed in the error propagation.

The comparison with the reduced-feedback slab-equilibrium calculation further constrains the physical interpretation of this branch. A homogeneous full-covering sandwich slab provides strong disc-corona radiative feedback and tends to over-cool the corona relative to the observed locus. Reproducing the corrected full-depth optical depths requires most of the local accretion power to be dissipated in the hot phase, $f\simeq1$, together with a feedback factor far below the full-covering value, $g\ll 1$. In this framework, $f$ measures the local partition of dissipated power between the disc and the corona, while $g$ should be regarded as an effective radiative-feedback parameter. The inferred $g\ll1$ therefore suggests weakened disc-corona feedback, as expected if the active regions are patchy.

The displacement of the R18 median point is also understandable in view of how its optical depth was obtained. R18 did not fit $\tau_{\rm T}$ directly for individual sources, but inferred it by calibrating the relation between $\Gamma$, $kT_{\rm e}$, and $\tau_{\rm T}$ using slab-\texttt{compPS} simulations over a broad grid extending to $\tau_{\rm T}=5.1$. This procedure is useful for identifying large-sample trends, but the absolute optical-depth normalization is sensitive to the assumed $E_{\rm C}$-to-$kT_{\rm e}$ conversion, the 2--10 keV fitting band, and the behaviour of \texttt{compPS} in the high-optical-depth slab regime. The absolute optical-depth scale may therefore be less secure in the high-$\tau_{\rm T}$ part of the calibration grid. We therefore regard the R18 median as a valuable external population-level reference, but not as a homogeneous measurement that can be combined directly with the source-by-source slab-\texttt{compTT} sample.

Multiple observational diagnostics indicate that the hot corona is a compact structure located in the inner region of the accretion disc \citep[e.g.,][]{2009ApJ...693..174C, 2015ApJ...806..251J, 2013MNRAS.431.2441D, 2014A&ARv..22...72U}. Therefore, a high local coronal dissipation fraction does not necessarily imply a strong conflict with observed broadband AGN spectra. The inner accretion disc, where most of the gravitational power is released, contributes primarily to the unobservable extreme-UV peak of the big blue bump in typical Seyfert galaxies \citep{1973A&A....24..337S}. If a large fraction of the inner-disc power is dissipated in the corona, the main effect would be to reduce the intrinsic thermal emission from the innermost disc and to enhance the Comptonized high-energy output. Because the peak of the inner-disc thermal component is usually hidden by absorption \citep{1999PASP..111....1K}, a corona-dominated local dissipation budget can be compatible with the observed optical to UV continuum, provided that the outer disc still supplies the observed lower-frequency thermal emission.

Our findings also prompt a re-examination of the widely discussed positive correlations between $\Gamma-\lambda_{\rm Edd}$ and $\kappa_{\rm 2-10~keV}-\lambda_{\rm Edd}$ \citep[e.g.,][]{2007MNRAS.381.1235V, 2008ApJ...682...81S, 2012MNRAS.425..623L, 2013MNRAS.433.2485B}. Conventionally, these trends are interpreted as evidence that the coronal dissipation fraction decreases as accretion rate increases. However, the toy model presented in Appendix \ref{sec:appendixD} qualitatively illustrates that both correlations can arise even when $f$ is held fixed. Therefore, the observed $\Gamma-\lambda_{\rm Edd}$ and $\kappa_{\rm 2-10~ keV}-\lambda_{\rm Edd}$ relations do not, by themselves, necessarily require a systematic variation of the disc-corona energy partition with accretion rate.

Correspondingly, we find no significant correlation between $kT_{\rm e}$ and $\lambda_{\rm Edd}$, or between $y$ and $\lambda_{\rm Edd}$, in the present \texttt{compTT} sample. This differs from the trend reported by R18 from the large \textit{Swift}/BAT sample. A likely reason is the systematic difference in sample selection and parameter estimation: the present sample is restricted to sources with direct slab-Comptonization measurements of $kT_{\rm e}$ and $\tau_{\rm T}$, whereas R18 relies on cutoff-based temperatures. In addition, hotter and optically thinner coronae produce stronger emission in the 14--195 keV BAT band and are therefore easier to detect, whereas cooler, optically thicker sources can fall below the BAT sensitivity threshold more easily, especially at low luminosities. The negative $kT_{\rm e}-\lambda_{\rm Edd}$ trend reported by R18 may therefore partly reflect selection effects and parameterisation differences inherent to a hard-X-ray flux-limited sample.

Although our main conclusions do not require a pure electron-positron pair plasma, pair balance remains a useful supplementary framework for understanding the thermal state of coronae. Previous studies in the compactness-temperature plane \citep[e.g.,][also see R18]{2015MNRAS.451.4375F} have noted that high-temperature AGN coronae may reside near the pair-regulation boundary, implying that pair production can act as a thermostat in some sources \citep[e.g.,][]{1982ApJ...258..335S, 1984ApJ...283..842Z, 1987ApJ...319..643L}. However, our supplementary calculations in Appendix \ref{sec:appendixE} suggest that the relatively low-temperature, high-optical-depth part of our sample is difficult to interpret as a purely thermal, pair-dominated homogeneous slab. In the pure-pair limit, such sources would require extremely large compactness values to supply the observed optical depths, which corresponds to the unusually large luminosity or small height of the corona. The fitted optical depths are therefore more naturally interpreted as being dominated by an electron--ion plasma, with pairs possibly contributing in some objects. If pairs are dynamically important in individual high-temperature sources, the plasma may need to be hybrid or non-thermal \citep[e.g.,][]{2015MNRAS.451.4375F}.

MCG--5--23--16 provides a useful example of a source for which the simple sandwich-slab interpretation may not apply straightforwardly. High-quality reflection modelling of this source suggests that the inner cold reflector can be truncated at several tens of gravitational radii, well outside the ISCO \citep{2023MNRAS.526.3540S}. Therefore, it may correspond to a recessed cold disc with an inner hot flow. In this sense, MCG--5--23--16 is a physically informative exception: its displacement from the main ridge is consistent with the idea that departures from the constant-$y$ locus can occur when the disc-corona geometry deviates from a sandwich-like slab corona configuration. More generally, identifying when and why the cold disc truncates in moderate-accretion Seyferts remains an important problem for future work \citep[e.g.,][]{2022MNRAS.515.2208G}.

Our results also raise an important physical question: what mechanism can maintain a high and relatively stable coronal dissipation fraction across AGNs with different accretion rates? In sandwich disc-corona models, two broad classes of coronal energy supply are usually considered. The first is thermal evaporation, where the hot coronal electrons evaporate the surface layer of the accretion disc through thermal conduction, forming an ADAF--SSD two-phase flow where gravitational energy is released during the inward accretion of the evaporated material \citep[e.g.,][]{2000A&A...361..175M, 2022iSci...25j3544L}. The second is magnetic buoyancy, where magnetic fields amplified by the MRI within the disc rise and dissipate in the corona \citep[e.g.,][]{2002MNRAS.332..165M, 2009MNRAS.394..207C}.

Both pictures face difficulties in explaining the high $f$ favoured by our results. In the thermal evaporation/ADAF scenario, the ADAF state is generally expected to remain stable only below a relatively low critical accretion rate ($\dot{m} \lesssim 0.01$). For luminous systems with $\dot{m} \gtrsim 0.1$, the coronal dissipation fraction sustained by such thermal flows is typically well below the level required here \citep[e.g.,][]{2022ApJ...932...97C}. For the standard magnetic buoyancy picture, the magnetic field strength and upward transport efficiency are constrained by the effective viscosity parameter $\alpha$ in the SSD, making it difficult to achieve the required large $f$ \citep[e.g.,][]{2002MNRAS.332..165M, 2025MNRAS.544.1748X}.

More generally, our results suggest that any viable disc-corona coupling mechanism should be able to channel a large fraction of the dissipated accretion power into the corona while remaining relatively insensitive to source-to-source variations. Mechanisms that enhance vertical magnetic energy transport in radiation-pressure-dominated inner discs, including photon-bubble-instability (PBI)-assisted buoyant transport, may therefore be promising directions for further theoretical study \citep[e.g.,][]{2026MNRAS.548ag599X}.

\section{Conclusions}

We have investigated the distribution of slab-corona measurements for radio-quiet Seyfert galaxies in the $kT_{\rm e}$--$\tau_{\rm T}$ plane using published direct slab-geometry thermal Comptonization measurements. Our main results are as follows.

First, motivated by the transport properties of bottom-illuminated slab coronae, we adopt a classically normalized effective parameter $y=4\theta\tau_{\rm T}K_3(1/\theta)/K_2(1/\theta)$. This form reflects the linear-$\tau_{\rm T}$ scaling for the slab corona while replacing the energy-gain factor with the Maxwell--Jüttner averaged one.

Second, after converting the slab-\texttt{compTT} optical depths to a common full-vertical-depth convention, the cleaned \texttt{compTT} sample $(N = 14)$ occupies a narrow anti-correlated ridge in the $kT_{\rm e}$--$\tau_{\rm T}$ plane. This ridge is well described by an approximately constant effective slab Comptonization parameter, with $\langle y\rangle=0.770$ and $\sigma_{\log y}=0.086$ dex.

Third, comparison with reduced-feedback slab-equilibrium calculations suggests that the observed locus is difficult to reproduce with low coronal dissipation fractions or with the homogeneous full-covering feedback limit. Within this framework, the observed locus favours a corona-dominated local energy budget, $f\simeq 1$, together with a feedback factor well below the full-covering value, $g\ll1$.

Finally, supplementary pair-balance estimates indicate that the low-temperature, high-optical-depth part of the sample is unlikely to be supplied by a purely thermal pair plasma alone. The fitted optical depths are therefore more plausibly dominated by electron--ion plasma, although pair regulation may still be relevant in some hotter sources.

Overall, our results suggest that Seyfert slab coronae occupy a narrow Comptonization branch. Within the reduced-feedback slab-equilibrium interpretation, this branch favours dominant local coronal dissipation and reduced disc--corona radiative feedback. This picture is consistent with patchy slab coronae, and provides a useful observational constraint on future models of vertical energy transport in accretion discs.

\section*{Acknowledgements} 

I am grateful to Prof. Xinwu Cao for his continuous support and helpful discussions throughout this work. I also thank Prof. Andrzej A. Zdziarski and Dr. Michał Szanecki for providing the convolution version of \texttt{compps}.

The author used ChatGPT (OpenAI) during manuscript preparation for language polishing, grammar checking, and assistance with table formatting and limited editorial suggestions. All scientific content, references, calculations, data selection, interpretation, and final wording were independently reviewed by the author, who assumes full responsibility for the accuracy of the manuscript.

Some of the simulations and analyses were carried out on the SilkRiver Supercomputer of Zhejiang University, located at the Zhejiang University Information Center. 

\section*{Data Availability}

The code and data underlying this work are available in the github repository at \url{https://github.com/XU-Haichao/code_used_in_my_constant_y_paper.git}. The repository contains the research scripts and data used in this study.



\bibliographystyle{mnras}
\bibliography{example} 




\appendix

\appendix

\section{List of the sources}
\label{sec:appendixA}

Here we provide source-by-source notes for the objects included in our radio-quiet Seyfert slab-corona sample. The main parent sample is taken from S24, who performed spectral and timing analyses of 46 \textit{NuSTAR} observations of 20 AGNs and reported best-fitting coronal temperatures and optical depths using the \texttt{compTT} model for both slab and spherical geometries. We adopt their slab-geometry results as the basis of the primary \texttt{compTT} dataset.

For the S24 parent sample, we revise the source classifications, redshifts, and Eddington ratios using the BASS DR2 catalogue whenever available, rather than adopting the values listed directly in S24. Unless otherwise specified, we use the optical spectroscopic AGN type, best redshift, and Eddington ratio reported in BASS DR2 \citep{2022ApJS..261....2K}, with the \textit{Swift}/BAT 70-month catalogue used as an additional cross-check for the BAT counterpart identification and hard-X-ray source classification \citep{2013ApJS..207...19B}.

To keep the sample focused on radio-quiet Seyfert coronae and to minimize possible jet-related contamination of the hard X-ray continuum, we exclude the two clearly radio-loud objects in the S24 sample, namely Mrk 6 and 4C 50.55. 

\textit{VLBA} observations reveal that Mrk 6 is a radio-loud Seyfert galaxy with a compact inverted-spectrum radio core and parsec-scale components resembling jet elements or hotspots \citep{2014MNRAS.440.2976K}. Its X-ray spectrum is affected by complex and variable line-of-sight absorption, including partial-covering and double partial-covering components, with a neutral absorber having column densities of order a few $10^{22}~\rm cm^{-2}$ and high covering fractions \citep{2024MNRAS.528.5269L}. In addition, X-ray spectroscopic studies have reported both ultrafast outflow and warm absorber components in this source \citep{2024MNRAS.528.4504K}. These properties make Mrk 6 unsuitable for defining a clean radio-quiet slab-corona sample.

Previous X-ray studies suggest that the hard X-ray emission of 4C 50.55 is not jet-dominated, but is likely dominated by an accretion-powered coronal continuum \citep{2010ApJ...721.1340T}; nevertheless, the source is a broad-line radio galaxy with strong radio emission and jet activity \citep{2007MNRAS.382..937M}. We therefore conservatively exclude it from the primary sample. 

In addition to the S24 sample, we include two radio-quiet Seyfert galaxies with high-quality published slab \texttt{compTT} measurements, Ark 120 and ESO 323-G77. We also compile a small set of slab \texttt{compPS} measurements for comparison, including NGC 4151, IC 4329A, and MCG-5-23-16. These \texttt{compPS} points are not used to define the primary \texttt{compTT} statistics, but are retained as a supplementary model-systematics check. The source-specific comments are summarized below.

\begin{itemize}
    \item 1H 0419-577 ($z=0.104$) is a luminous radio-quiet Seyfert 1 galaxy with strong long-term spectral variability, a prominent soft excess, variable absorption/reflection signatures \citep[e.g.,][]{2002MNRAS.330L...1P, 2013MNRAS.435.1287P, 2018MNRAS.473.3584P}. The 2015 \textit{NuSTAR}/\textit{Swift} observation supported a low-temperature thermal Comptonization corona \citep{2018MNRAS.476.1258T}. We adopt the S24 joint analysis based on three \textit{NuSTAR} epochs obtained on 2015 June 3, 2018 May 15, and 2018 November 13. S24 fitted these epochs with an absorbed Comptonized continuum plus non-relativistic reflection and allowed the intrinsic absorption column to vary between observations.
    
    \item Ark 564 ($z=0.0243$) is a radio-quiet narrow-line Seyfert 1 galaxy with previous \textit{XMM--Newton} studies showing strong soft-excess emission and rapid continuum variability \citep[e.g.,][]{2004MNRAS.347..854V}. We adopt the S24 joint analysis based on three \textit{NuSTAR} epochs obtained on 2015 May 22, 2018 June 9, and 2018 November 28. Its Eddington ratio was computed using the reverberation-mapping black-hole mass from \citet{2004ApJ...613..682P} and a bolometric correction applied to the X-ray luminosity.
    
    \item ESO 103-G35 ($z=0.0135$) is a nearby obscured Seyfert galaxy, optically classified as a Sy 1.9 and often treated as a Sy 2 in hard-X-ray studies \citep[e.g.,][]{1979A&A....76L..14P, 2001MNRAS.324..521A}. Its X-ray spectrum is characterized by substantial line-of-sight absorption, narrow Fe K$\alpha$ emission, and reflection/absorption complexity \citep{2018MNRAS.481.4419B}. We adopt the S24 joint \textit{NuSTAR} analysis based on two epochs obtained on 2013 February 24 and 2017 October 15. 
    
    \item ESO 362-G18 ($z=0.0125$) is a Seyfert 1.5 galaxy whose X-ray spectrum generally resembles that of a type-1 AGN, showing a prominent soft excess and reflection features \citep{2014MNRAS.443.2862A, 2021ApJ...913...13X}, although multi-epoch observations revealed variable neutral absorption \citep{2014MNRAS.443.2862A}. We adopt the S24 \textit{NuSTAR} analysis based on the epoch obtained on 2016 September 24.

    \item ESO 383-G18 ($z=0.0129$) is a nearby Seyfert 2 galaxy with a strongly absorbed X-ray spectrum. Previous broadband X-ray studies modelled its spectrum with multiple cold, partially covering absorbers, consistent with a clumpy torus plus additional Compton-thin absorbing material along the line of sight \citep{2010A&A...518A..47R}. We adopt the S24 joint analysis based on two \textit{NuSTAR} epochs obtained on 2014 September 11 and 2016 January 20.

    \item GRS 1734-292 ($z=0.0218$) is a nearby Seyfert 1 galaxy with moderate absorption and reflection features, together with evidence for a low high-energy cutoff \citep{2017MNRAS.466.4193T}. Although a compact double-sided radio structure has been reported, its radio and X-ray luminosities are consistent with a radio-quiet AGN classification \citep{2017MNRAS.466.4193T}. We adopt the S24 joint \textit{NuSTAR} analysis based on two epochs obtained on 2014 September 16 and 2018 May 28.

    \item HE 1143-1810 ($z=0.0326$) is a bare Seyfert 1 galaxy with no significant intrinsic neutral or warm absorption detected \citep{2011A&A...530A.125C}. A joint \textit{XMM--Newton}/\textit{NuSTAR} monitoring campaign in December 2017 revealed a UV--X-ray correlation, a soft excess, and a high-energy cutoff implying a comparatively low electron temperature for the hot corona \citep{2020A&A...634A..92U}. We adopt the S24 joint \textit{NuSTAR} analysis based on five epochs obtained on 2017 December 16, 18, 20, 22, and 24.

    \item IC 4329A ($z=0.0160$) is a bright, nearby radio-quiet Seyfert 1.2 galaxy whose 2012 simultaneous \textit{NuSTAR}/\textit{Suzaku} spectrum revealed modest continuum variability, mild absorption, Compton reflection, and a well-constrained high-energy cutoff, enabling detailed modelling of the hot corona \citep{2014ApJ...781...83B, 2014ApJ...788...61B}. We adopt the S24 joint \textit{NuSTAR} slab-\texttt{compTT} analysis of the same campaign, based on the \textit{NuSTAR} observation obtained on 2012 August 12. Since IC 4329A also has a published slab-\texttt{compPS} measurement from the same 2012 campaign, we include that result in the separate \texttt{compPS} comparison sample \citep{2014ApJ...781...83B}.

    \item MCG-5-23-16 ($z=0.00841$) is a bright Seyfert 1.9 galaxy with moderate absorption and a hard X-ray spectrum in which the primary-continuum curvature can be separated from the Compton-reflection curvature \citep{2015ApJ...800...62B}. A key feature of MCG-5-23-16 is that its broad Fe K$\alpha$ reflection component indicates a truncated inner accretion disc at tens of gravitational radii ($R_{\rm g}$). The recent simultaneous \textit{XMM--Newton}/\textit{NuSTAR} analysis found $R_{\rm in}=40^{+23}_{-16}R_{\rm g}$, broadly consistent with earlier reflection studies, although the inferred radius is partly degenerate with the disc inclination \citep{2023MNRAS.526.3540S}. We adopt the S24 joint slab-\texttt{compTT} analysis based on five \textit{NuSTAR} epochs obtained on 2012 July 11, 2013 June 3, 2015 February 15, 2015 February 21, and 2015 March 13. Since MCG-5-23-16 also has published slab-\texttt{compPS} measurements, we include those results in the separate \texttt{compPS} comparison sample \citep{2015ApJ...800...62B, 2022MNRAS.516.5907M, 2023MNRAS.526.3540S}. The 2012--2013 slab-\texttt{compPS} measurement overlaps with the epochs used in S24, and we therefore assign it the same S24/BASS Eddington ratio for consistency. For the independent 2022 slab-\texttt{compPS} measurement, we use an epoch-specific $\lambda_{\rm Edd}\simeq0.038$ estimated from the 2022 X-ray luminosity on the same bolometric-correction scale.

    \item MCG+8-11-11 ($z=0.0202$) is a bright, radio-quiet, unobscured Seyfert 1 galaxy. Earlier broad-band X-ray studies found a hard X-ray continuum with warm absorption, Fe K$\alpha$ emission, and Compton/reflection complexity, while the dedicated \textit{NuSTAR} analysis detected a high-energy cutoff at $\sim150-200$ keV together with a weak relativistic Fe K$\alpha$ component and disc reflection \citep{2010A&A...522A..64B, 2018MNRAS.473.3104T}. We adopt the S24 joint slab-\texttt{compTT} analysis based on the \textit{NuSTAR} epoch obtained on 2016 August 16.

    \item Mrk 110 ($z=0.0352$) is a bright, historically radio-quiet Seyfert 1 galaxy with no significant intrinsic X-ray absorption and a nearly face-on view of the disc--corona system. Broad-band \textit{XMM--Newton}/\textit{NuSTAR} studies show an absorption-free soft excess, weak reflection hump, and a continuum well described by warm and hot Comptonization plus mild disc reflection \citep{2021A&A...654A..89P}. We adopt the S24 joint slab-\texttt{compTT} analysis based on the \textit{NuSTAR} epochs obtained on 2017 January 23, 2019 November 16, and 2020 April 5. We retain Mrk 110 in the main sample, but note that recent VLBI observations have revealed intermittent relativistic jet activity in this historically radio-quiet source \citep{2025ApJ...987L..26W}.

    \item Mrk 509 ($z=0.0347$) is a bright radio-quiet Seyfert 1.2 galaxy. Previous multiwavelength campaigns showed a prominent soft X-ray excess and complex optical/UV--X-ray variability, with warm Comptonization often invoked for the soft-excess component \citep[e.g.,][]{2011A&A...534A..39M, 2013A&A...549A..73P}. The \textit{NuSTAR}/\textit{Suzaku} observations of Mrk 509 showed a power-law dominated hard X-ray continuum together with reflection signatures and a soft-excess/reflection modelling degeneracy \citep{2019ApJ...871...88G}. We adopt the S24 joint slab-\texttt{compTT} analysis based on two \textit{NuSTAR} epochs obtained on 2015 April 29 and 2015 June 2.

    \item NGC 3281 ($z=0.0111$) is a nearby radio-quiet Seyfert 2 galaxy with a heavily obscured nucleus. Earlier broad-band X-ray and mid-infrared studies identified it as a Compton-thick source and associated the obscuration with a dusty, clumpy torus, together with prominent Fe K$\alpha$ emission \citep{2002A&A...381..834V, 2011ApJ...738..109S}. In S24, the \textit{NuSTAR} spectra are treated as a severely absorbed source and modelled with an absorbed Comptonized continuum plus neutral torus reflection, allowing the line-of-sight absorption to vary between epochs. We adopt the S24 joint slab-\texttt{compTT} analysis based on two \textit{NuSTAR} epochs obtained on 2016 January 22 and 2020 July 15.

    \item NGC 5506 ($z=0.00598$) is catalogued as a Seyfert 1.9 in the BASS/BAT catalogues, but is commonly interpreted as a bright Compton-thin, optically obscured narrow-line Seyfert 1 galaxy, whose near-infrared spectrum reveals direct evidence for a reddened broad-line region \citep{2002A&A...391L..21N}. The \textit{NuSTAR}/\textit{Swift} spectrum is well described by an absorbed continuum, distant reflection, and narrow ionized iron lines, with no strong requirement for relativistically blurred reflection \citep{2015MNRAS.447.3029M}. We adopt the S24 joint slab-\texttt{compTT} analysis of the same \textit{NuSTAR} observation obtained on 2014 April 1. 

    \item NGC 5728 ($z=0.0103$) is a nearby Seyfert 1.9 galaxy with a heavily obscured nucleus and well-known ionization cones. Multiwavelength studies show a complex circumnuclear environment, including AGN-driven outflows, a star-forming ring, and extended X-ray emission associated with photoionized and shocked gas \citep[e.g.,][]{2019MNRAS.490.5860S, 2023ApJ...950..143T}. In S24, its \textit{NuSTAR} spectrum is treated as a typical absorbed hard-X-ray source and modelled with an absorbed Comptonized continuum plus distant neutral reflection. We adopt the S24 joint slab-\texttt{compTT} analysis based on two \textit{NuSTAR} epochs obtained on 2013 January 2 and 2020 July 13.

    \item NGC 6814 ($z=0.00579$) is a nearby type-1 Seyfert galaxy with strong X-ray and UV/optical variability \citep[e.g.,][]{2024MNRAS.527.5569G}. Previous \textit{NuSTAR} spectroscopy found a high-energy cutoff, a slightly broadened Fe K$\alpha$ line with associated relativistic disc reflection, and a narrow Fe K$\alpha$ component likely produced in distant Compton-thin material \citep{2018MNRAS.473.3104T}. We adopt the S24 joint slab-\texttt{compTT} analysis based on the \textit{NuSTAR} epoch obtained on 2016 July 4, where the spectrum required both distant and relativistic reflection components but no intrinsic cold absorption.

    \item SWIFT J2127.4+5654 ($z=0.0149$) is a narrow-line Seyfert 1 galaxy with previous studies reporting a relativistically broadened Fe K$\alpha$ line and an intermediate black-hole spin \citep{2009MNRAS.398..255M, 2014MNRAS.440.2347M}. The simultaneous 2012 November \textit{XMM--Newton}/\textit{NuSTAR} campaign covered the 0.5--80 keV band and found a steep intrinsic continuum together with a well-constrained high-energy cutoff ($E_{\rm c}=108^{+11}_{-10}$ keV) \citep{2014MNRAS.440.2347M}. We adopt the S24 joint slab-\texttt{compTT} analysis based on the four \textit{NuSTAR} epochs obtained on 2012 November 4, 5, 6, and 8.

    \item UGC 6728 ($z=0.00597$) is a nearby low-mass Seyfert 1 galaxy, also described as a bare AGN with little line-of-sight obscuration \citep{2024MNRAS.532.1185N}. Its hard X-ray emission is therefore plausibly associated with the inner accretion flow and Comptonizing plasma rather than absorption variability. We adopt the S24 joint slab-\texttt{compTT} analysis based on the two \textit{NuSTAR} epochs obtained on 2016 July 10 and 2017 October 13. We retain UGC 6728 in the main sample, but treat it as a borderline case because it was removed as a possible star-forming-galaxy contaminant in a later BASS/\textit{Fermi} sample selection based on SFG catalogue and BPT diagnostics \citep{2025NatAs...9.1086L}.

    \item Ark 120 ($z=0.0326$) is a radio-quiet, bare Seyfert 1 nucleus with little or no intrinsic absorption along the line of sight \citep{2004MNRAS.351..193V,2021MNRAS.506.3111N}. Deep \textit{XMM-Newton}/\textit{NuSTAR} observations show a prominent soft excess and Fe K$\alpha$ complex, with the broadband continuum well described by warm and hot Comptonization plus mildly relativistic disc reflection \citep{2018A&A...609A..42P}. We adopt the slab-\texttt{compTT} analysis of \citet{2019A&A...623A..12M} based on the simultaneous \textit{XMM-Newton}/\textit{NuSTAR} epoch obtained on 2014 March 22, for which both the hot-corona temperature and optical depth are constrained; the 2013 February 18 observation is not used because the slab-\texttt{compTT} fit gives only a lower limit on $kT_{\rm e}$.

    \item ESO 323-G77 ($z=0.0156$) is a nearby polar-scattered Seyfert 1.2 galaxy and X-ray changing-look AGN with strong X-ray absorption variability, previously observed in unobscured, Compton-thin, and Compton-thick states \citep{2014MNRAS.437.1776M,2016MNRAS.457..510S,2023A&A...672A..10S}. We adopt the joint slab-\texttt{compTT} analysis of \citet{2023A&A...672A..10S} based on the five \textit{NuSTAR} epochs obtained on 2016 December 14, 2016 December 20, 2017 January 4, 2017 February 3, and 2017 March 31, supplemented by the combined \textit{Swift}/XRT spectrum. During this campaign the source was caught in a persistent Compton-thin obscured state, and the \texttt{MYTorus} model with a slab-\texttt{compTT} continuum provided finite constraints on the hot-corona temperature and optical depth.

    \item NGC 4151 ($z=0.00332$) is a nearby Seyfert 1.5 galaxy that is securely radio-quiet according to its radio-to-X-ray ratio, although VLBA observations reveal a weak two-sided parsec-scale radio jet \citep{2005AJ....130..936U}. We adopt the joint slab-\texttt{compPS} analysis of \citet{2010MNRAS.408.1851L}, who fitted the extreme bright and dim hard-X-ray states using accumulated \textit{INTEGRAL} spectra together with contemporaneous \textit{RXTE}, \textit{XMM-Newton}, and \textit{Suzaku} data. The bright-state spectrum is based on the 2003 May observations, including \textit{INTEGRAL} revolutions 0074--0075 and contemporaneous \textit{RXTE}/\textit{XMM-Newton} coverage, whereas the dim-state spectrum combines low-flux observations from 2005 April to 2007 May. Since these are state-resolved measurements rather than a single source-averaged measurement, we assign state-dependent Eddington ratios, $\lambda_{\rm Edd}=0.019$ and $0.010$ for the bright and dim states, respectively, from the bolometric estimates of \citet{2010MNRAS.408.1851L}, instead of adopting a single BASS catalogue value.

\end{itemize}

\section{Radiative-transfer quantities for the sandwich disc-corona model}
\label{sec:appendixB}

In this appendix we show the radiative-transfer quantities entering the sandwich disc-corona model as functions of the coronal temperature $kT_{\rm e}$ and full vertical optical depth $\tau_{\rm T}$. These quantities include the downward fraction of the Comptonized luminosity $\eta$, the Compton amplification factor $A$, the scattering probability of disc photons in the corona $p_{\rm sc}$, and the effective energy-integrated albedo $a$. We compute the Comptonization process using the scattering-order-resolved \texttt{compPSc} calculation. The downward Comptonized spectrum is then used as the incident spectrum for the convolution model \texttt{ireflect}, from which we obtain the Compton reflection hump and the corresponding effective albedo. 

\begin{figure*}
\includegraphics[width=0.8\linewidth]{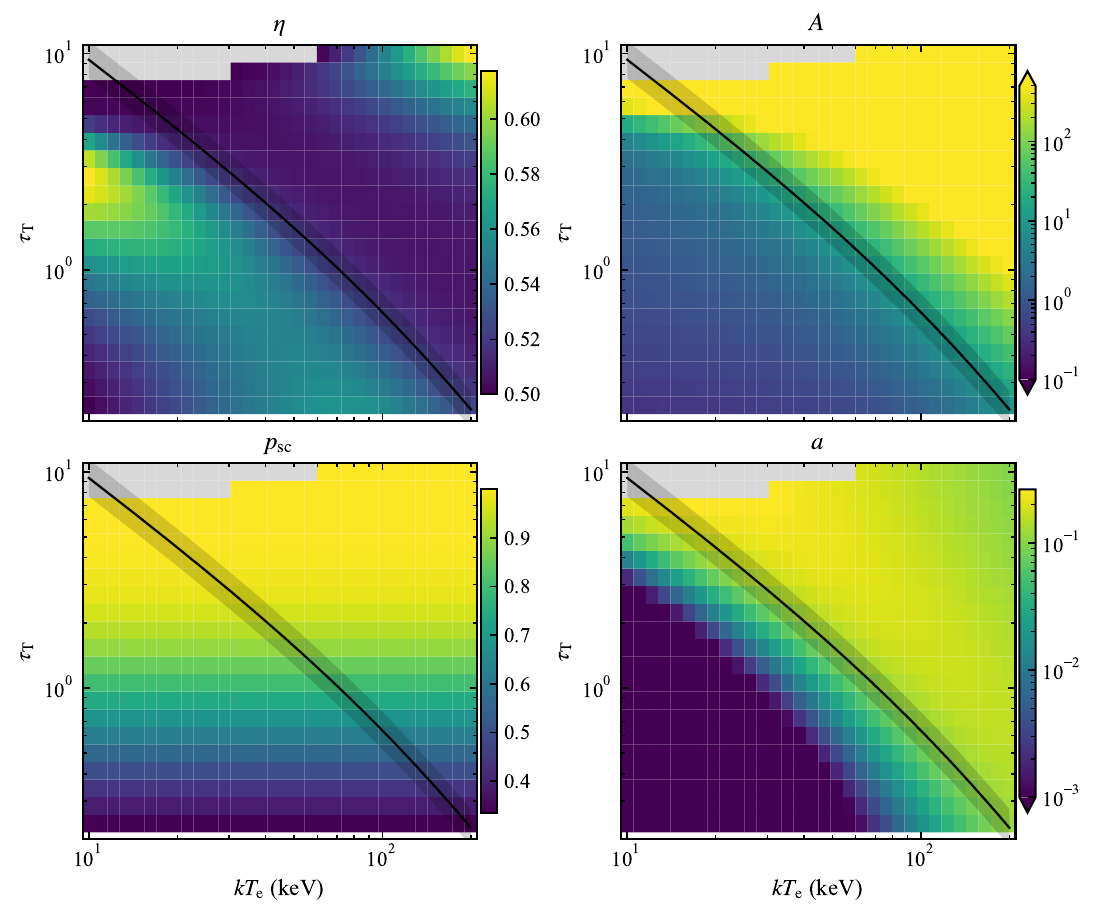}
\caption{Maps of radiative-transfer quantities. For each pair of electron temperature $kT_{\rm e}$ and vertical Thomson optical depth $\tau_{\rm T}$, we use the \texttt{compPSc} calculation to compute the downward fraction of the Comptonized luminosity $\eta$, the Compton amplification factor $A$, and the scattering probability $p_{\rm sc}$. The effective energy-integrated albedo $a$ is computed by applying \texttt{ireflect} to the downward Comptonized spectrum. Grey regions denote grid points for which the calculation did not converge. The black curve shows the empirical constant-$y$ ridge of the cleaned \texttt{compTT} sample. The shaded band indicates the corresponding logarithmic scatter.}
\label{fig:parameter_grid}
\end{figure*}

We approximate the local disc emission by a blackbody with $kT_{\rm bb}=5~{\rm eV}$. Motivated by the empirical relation between Eddington ratio and disc ionization obtained by \citet{2011ApJ...734..112B}, we adopt an ionization parameter $\xi=100~{\rm erg~cm~s^{-1}}$ for the reflecting disc surface. We fix metal and iron abundances to solar values \citep{1989GeCoA..53..197A}. We verified that varying the adopted blackbody temperature and ionization parameter over representative values does not appreciably change the resulting equilibrium boundaries. In the \texttt{compPSc} calculation, the maximum scattering order is adjusted according to the optical depth and is allowed to reach 2000 for the large-$\tau_{\rm T}$ cases, in order to ensure convergence in the mildly optically thick regime. The resulting maps of $\eta, A, p_{\rm sc}$ and $a$ across the $kT_{\rm e}-\tau_{\rm T}$ plane are shown in Figure \ref{fig:parameter_grid}.

The four quantities show different dependences across the parameter space explored here. The downward fraction $\eta$ varies only mildly, remaining slightly larger than 0.5 and reaching 0.6 in part of the grid, which is consistent with the calculations in HM91 and HM93. This behaviour is physically expected. At high scattering orders, the radiation field approaches the diffusion limit, and the upper and lower slab surfaces become nearly symmetric escape boundaries \citep[e.g.,][]{1980A&A....86..121S,2026arXiv260601087X}. At low scattering orders, however, the seed photons are initially injected upward from the disc surface. Photons that are redirected downward after scattering can gain somewhat more energy, leading to a modest excess of downward Comptonized luminosity \citep[e.g.,][also see HM91, HM93]{2000ApJ...540..131P, 2001ApJ...556..716P}. Since the low-order components do not dominate the bolometric Comptonized output over most of the relevant parameter space, this anisotropy does not drive $\eta$ far from 1/2. 

The scattering probability $p_{\rm sc}$ is controlled mainly by optical depth and depends only weakly on $kT_{\rm e}$. This is consistent with the fact that the first scattering of the disc blackbody photons occurs deep in the Thomson regime for the temperatures considered here. In the present calculation the seed photons are injected normally from the lower boundary, for which the simple Thomson estimate $p_{\rm sc}= 1-\exp(-\tau_{\rm T})$ captures the monotonic increase of $p_{\rm sc}$ with optical depth seen in Figure \ref{fig:parameter_grid}. For a Lambert-law angular distribution at the lower boundary, angle-averaging the no-scattering probability gives the corresponding single-crossing estimate $p_{\rm sc}=1-2E_3(\tau_{\rm T})$, where $E_3$ is the exponential integral.

In contrast, the amplification factor $A$ varies by orders of magnitude. It is small in the low-temperature, low-optical-depth regime, where photons undergo few scatterings and gain little energy per scattering, but increases rapidly as both $kT_{\rm e}$ and $\tau_{\rm T}$ increase. The effective albedo $a$ is negligible where the incident spectrum contains little hard X-ray power, and becomes appreciable when a substantial fraction of the downward Comptonized luminosity emerges in the hard X-ray band where Compton reflection is efficient. 

Along the empirical constant-$y$ ridge traced by the observed sources, $\eta$ and $a$ vary more slowly than $A$; representative values $\eta\simeq0.5-0.6$ and $a\simeq0.1-0.2$ provide useful order-of-magnitude estimates, which is also consistent with the calculations in HM91. By contrast, $A$ shows a much stronger dependence on position in the $kT_{\rm e}-\tau_{\rm T}$ plane, from values below $\sim10$ at the high-temperature, low-$\tau_{\rm T}$ end to values above $\sim100$ toward the low-temperature, high-$\tau_{\rm T}$ end. Therefore, detailed equilibrium calculations should use the radiative-transfer value of $A$, rather than a simple analytic approximation.

\section{Posterior calculation, intrinsic scatter, and covariance tests}
\label{sec:appendixC}

In this appendix we describe how the posterior distribution in the $f-g$ plane is computed and test its sensitivity to the assumed intrinsic scatter $\sigma_{\rm int}$ and to the unknown covariance between the published $kT_{\rm e}$ and $\tau_{\rm T}$ measurements. The purpose of this calculation is not to identify a single best-fitting parameter set, but to identify the region in the $f-g$ plane within which the model can reproduce the main ridge in the $kT_{\rm e}-\tau_{\rm T}$ plane.

For each pair of global parameters $(f,g)$, the sandwich disc-corona model predicts a radiative-equilibrium curve in the $kT_{\rm e}-\tau_{\rm T}$ plane. In our calculation, we use the radiative-transfer quantities tabulated in Appendix \ref{sec:appendixB}, namely $A^{\rm cal}$, $\eta^{\rm cal}$, $p_{\rm sc}^{\rm cal}$, and $a^{\rm cal}$, and compute the Compton amplification required by the two-phase energy balance as
\begin{equation}
A^{\rm req}(kT_{\rm e},\tau_{\rm T};f,g)=
\frac{1+d\,p_{\rm sc}^{\rm cal}}
{g(1-a^{\rm cal})\eta^{\rm cal}
+d\left(1-g a^{\rm cal}\eta^{\rm cal}p_{\rm sc}^{\rm cal}\right)},
\label{eq:appc_areq}
\end{equation}
where $d=(1-f)/f$. The residual used to define the energy-balance condition in logarithmic space is
\begin{equation}
R(kT_{\rm e},\tau_{\rm T};f,g)
=
\ln\left[
\frac{A^{\rm cal}(kT_{\rm e},\tau_{\rm T})}
{A^{\rm req}(kT_{\rm e},\tau_{\rm T};f,g)}
\right].
\label{eq:appc_residual}
\end{equation}

Because both $kT_{\rm e}$ and $\tau_{\rm T}$ are positive quantities and span different orders of magnitude, the likelihood is constructed in logarithmic space. For the $i$-th source, the observational uncertainties are transformed into logarithmic space. For the electron temperature, the lower and upper log-space errors are
\begin{equation}
\sigma^-_{kT_{\rm e},i}
=
\ln\left(\frac{kT_{{\rm e},i}}{kT_{{\rm e},i}^-}\right),
\qquad
\sigma^+_{kT_{\rm e},i}
=
\ln\left(\frac{kT_{{\rm e},i}^+}{kT_{{\rm e},i}}\right),
\end{equation}
where $kT_{{\rm e},i}^-$ and $kT_{{\rm e},i}^+$ are the lower and upper bounds implied by the quoted uncertainty. The corresponding quantities for the optical depth are
\begin{equation}
\sigma^-_{\tau_{\rm T},i}
=
\ln\left(\frac{\tau_{{\rm T},i}}{\tau_{{\rm T},-,i}}\right),
\qquad
\sigma^+_{\tau_{\rm T},i}
=
\ln\left(\frac{\tau_{{\rm T},+,i}}{\tau_{{\rm T},i}}\right).
\end{equation}

In the fiducial calculation, we use symmetrized log-space errors,
\begin{equation}
\sigma_{kT_{\rm e},i,{\rm obs}}
=
\frac{1}{2}\left(\sigma^+_{kT_{\rm e},i}+\sigma^-_{kT_{\rm e},i}\right),
\qquad
    \sigma_{\tau_{\rm T},i,{\rm obs}}
=
\frac{1}{2}\left(\sigma^+_{\tau_{\rm T},i}+\sigma^-_{\tau_{\rm T},i}\right).
\end{equation}
We tested that retaining the asymmetric errors instead of using the symmetrised values does not significantly alter the posterior distribution in the $f-g$ plane.

We also include an additional isotropic intrinsic scatter term $\sigma_{\rm int}$ in the logarithmic $kT_{\rm e}-\tau_{\rm T}$ plane. This term prevents sources with very small quoted statistical errors from dominating the fit and phenomenologically accounts for source-to-source diversity and residual model systematics. When $\sigma_{\rm int}$ increases, the relative weights of sources with small and large statistical errors become more similar, and the posterior distribution in the $f-g$ plane broadens. Since we introduce $\sigma_{\rm int}$ as an additional error, increasing $\sigma_{\rm int}$ will lead to the dispersion of the posterior distribution in the $f-g$ plane. In the calculation of Figure \ref{fig:model}, $\sigma_{\rm int}$ is taken to be 0.25.

The theoretical constraint is an implicit curve $R(kT_{\rm e}, \tau_{\rm T};f,g)=0$, not a single-valued function with an error-free independent variable. We therefore construct the likelihood from the local distance of each observed point to this implicit curve. Around the observed point $(kT_{{\rm e},i},\tau_{{\rm T},i})$, we linearize the residual as
\begin{equation}
R(kT_{\rm e}, \tau_{\rm T};f,g)
\simeq
R_i+\left.\frac{\partial R}{\partial \ln kT_{\rm e}}\right|_i\ln\left(\frac{kT_{\rm e}}{kT_{{\rm e},i}}\right) + \left.\frac{\partial R}{\partial \ln\tau_{\rm T}}\right|_i\ln\left(\frac{\tau_{\rm T}}{\tau_{{\rm T},i}}\right),
\end{equation}
where
\begin{equation}
R_i=R(kT_{{\rm e},i},\tau_{{\rm T},i};f,g).
\end{equation}

The source papers generally provide one-dimensional confidence intervals for $kT_{\rm e}$ and $\tau_{\rm T}$. Although some individual source studies present two-dimensional confidence contours, the full numerical covariance information is not available in a homogeneous form for the whole sample. The fiducial calculation therefore adopts independent logarithmic observational errors, which should be understood as an error-model approximation, not as a physical assumption that the fitted Comptonization parameters are intrinsically independent. 

To test the possible impact of the known $kT_{\rm e}-\tau$ degeneracy in thermal Comptonization fits, we also allow the logarithmic observational errors to have a representative correlation coefficient $\rho_{T_{\rm e}\tau}$. In this case, the observational uncertainty is 
\begin{equation}
\begin{aligned}
    \sigma_{R,i,{\rm obs}}^2
=&\left(\left.\frac{\partial R}{\partial \ln kT_{\rm e}}\right|_i\sigma_{kT_{\rm e},i,{\rm obs}}\right)^2
+
\left(\left.\frac{\partial R}{\partial \ln\tau_{\rm T}}\right|_i \sigma_{\tau_{\rm T},i,{\rm obs}}\right)^2\\
&+
2\rho_{T_{\rm e}\tau}
\left.\frac{\partial R}{\partial \ln kT_{\rm e}}\right|_i\left.\frac{\partial R}{\partial \ln\tau_{\rm T}}\right|_i
\sigma_{kT_{\rm e},i,{\rm obs}}\,
\sigma_{\tau_{\rm T},i,{\rm obs}},
\end{aligned}
\label{eq:appc_sigmar_obs}
\end{equation}
and the total uncertainty projected onto the implicit residual is 
\begin{equation}
\sigma_{R,i}^2
=\sigma_{R,i,{\rm obs}}^2+
\left(\left.\frac{\partial R}{\partial \ln kT_{\rm e}}\right|_i^2+\left.\frac{\partial R}{\partial \ln\tau_{\rm T}}\right|_i^2\right)\sigma_{\rm int}^2.
\label{eq:appc_sigmar}
\end{equation}
The correlation coefficient $\rho_{T_{\rm e}\tau}$ is applied only to the published observational uncertainties, because it represents the possible covariance induced by spectral-fitting degeneracy. By contrast, $\sigma_{\rm int}$ is introduced as an isotropic phenomenological broadening, accounting for source-to-source diversity and residual model systematics rather than measurement covariance. We therefore add the intrinsic term without an additional off-diagonal covariance component.

Under the local linear approximation, this construction corresponds to projecting the two-dimensional measurement uncertainty onto the normal direction of the implicit curve $R=0$. It is therefore more appropriate for the present problem than a standard vertical-residual $\chi^2$, because neither $kT_{\rm e}$ nor $\tau_{\rm T}$ is treated as an error-free coordinate \citep[e.g.,][]{2010arXiv1008.4686H}.

Since $\sigma_{R,i}$ depends on $(f,g)$ through the local gradient of $R$, the Gaussian normalization is not a constant. We therefore use the full log-likelihood
\begin{equation}
\ln {\cal L}(f,g)
=
-\frac{1}{2}
\sum_i
\left[
\chi_i^2(f,g)
+
\ln\left(2\pi\sigma_{R,i}^2(f,g)\right)
\right],
\label{eq:appc_likelihood}
\end{equation}
where
\begin{equation}
\chi_i^2(f,g)=
\frac{R_i^2(f,g)}{\sigma_{R,i}^2(f,g)} .
\label{eq:appc_chi2}
\end{equation}

\begin{figure*}
\includegraphics{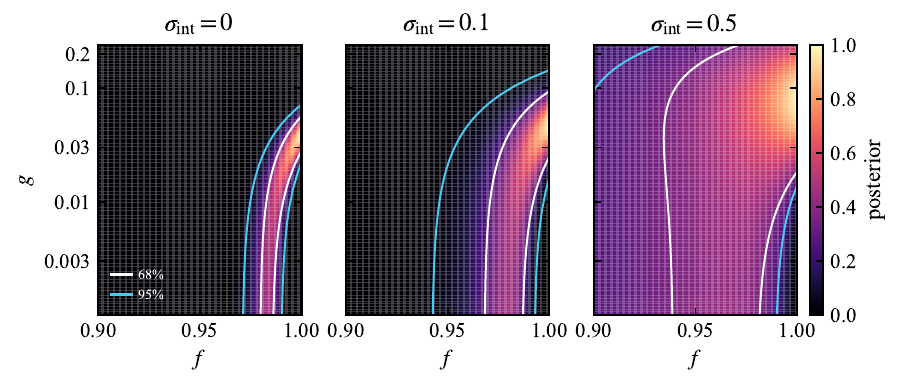}
\caption{Posterior distribution in the $f-g$ plane for different assumed intrinsic scatters. Each panel is obtained by fitting the cleaned \texttt{compTT} sample with the sandwich disc-corona model. The three panels adopt $\sigma_{\rm int}=0$, 0.1, and 0.5, respectively. White and cyan contours mark the 68\% and 95\% credible regions, respectively. Increasing $\sigma_{\rm int}$ broadens the posterior and reduces the dominance of the most tightly constrained individual sources.}
\label{fig:posterior}
\end{figure*}

A broad preliminary search over the $f-g$ plane shows that low-$f$ and large-$g$ solutions are strongly disfavoured, as is also evident from the equilibrium-boundary grid shown in Figure \ref{fig:model}. We therefore present a local posterior analysis over $0.9\le f\le 1$ and $10^{-3}\le g\le 0.24$, with uniform priors in $f$ and $g$ within this restricted domain. Although $g$ is displayed on a logarithmic axis, the credible contours are computed by integrating the posterior density over the linear cell area. The fiducial posterior shown in Figure \ref{fig:model} adopts $\sigma_{\rm int}=0.25$. To test the sensitivity of the inference to this assumption, Figure \ref{fig:posterior} repeats the posterior calculation for several values of $\sigma_{\rm int}$. The fiducial value used in Figure \ref{fig:model}, $\sigma_{\rm int}=0.25$, lies between the cases shown here. As expected, increasing $\sigma_{\rm int}$ broadens the posterior. However, the qualitative conclusion is unchanged. The posterior remains concentrated toward high local coronal dissipation, $f\simeq1$, together with substantially reduced feedback factors, $g\ll1$.

We also tested whether the posterior depends on the fiducial assumption $\rho_{T_{\rm e}\tau}=0$. Keeping the intrinsic logarithmic broadening fixed at $\sigma_{\rm int}=0.25$, we repeated the posterior calculation with representative negative correlations, $\rho_{T_{\rm e}\tau}=-0.5$ and $-0.9$. The resulting marginal constraints are nearly unchanged: the posterior remains concentrated toward $f$ close to the upper boundary of the explored range and low feedback factors $g$. The credible regions become only slightly narrower as the imposed anti-correlation is strengthened. Thus, within the adopted intrinsic-scatter model, the inferred preference for a high-dissipation, weak-feedback branch is not an artefact of setting the observational $kT_{\rm e}-\tau_{\rm T}$ covariance to zero.

\section{A toy model for the correlation of \texorpdfstring{$\Gamma-\lambda_{\rm Edd}$}{Gamma-mdot} and \texorpdfstring{$\kappa_{2-10{\rm keV}}-\lambda_{\rm Edd}$}{kappa-mdot}}
\label{sec:appendixD}

Here we present a simple toy model showing that the observed $\Gamma-\lambda_{\rm Edd}$ and $\kappa_{2-10{\rm keV}}-\lambda_{\rm Edd}$ trends can arise even if the underlying disc-corona parameters do not vary systematically with accretion rate. These correlations therefore do not, by themselves, imply that $f$ or $g$ must change with increasing $\dot{m}$.

\subsection{The correlation between \texorpdfstring{$\Gamma$}{Gamma} and \texorpdfstring{$\lambda_{\rm Edd}$}{mdot}}

As a phenomenological approximation, we represent the upward-escaping Comptonized spectrum by a cutoff power law,
\begin{equation}
     N(E) = N_0 E^{-\Gamma} \exp(-E/E_{\rm c}),\quad E\gtrsim E_{\rm s},
     \label{eq:ctofpl}
\end{equation}
where $E_{\rm c}$ is the high-energy cutoff and $E_{\rm s}$ marks the low-energy turnover. In rigorous thermal Comptonization theory, the photon index $\Gamma$ is determined by $kT_{\rm e}$, $\tau_{\rm T}$, and the effective Compton $y$-parameter.

For a standard SSD, the effective temperature of the thermal disc emission scales with the dimensionless accretion rate $\dot{m}$ as:
\begin{equation}
     kT_{\rm s} \propto \dot{m}^{1/4}
\end{equation}
For a blackbody spectrum, the photon energy at the spectral peak is approximately $2.82 k T_{\rm s}$. In a cutoff power-law description of the Comptonized spectrum, the low-energy turnover is set by seed photons that have undergone only a small number of scatterings. Since such photons retain an energy scale proportional to that of the original disc photons, the turnover energy satisfies
\begin{equation}
     E_{\rm s}\propto 2.82kT_{\rm s} \propto \dot{m}^{1/4}.
\end{equation}
Correspondingly, the photon number density near the turnover is proportional to that near the peak of the disc emission:
\begin{equation}
     N(E_{\rm s}) \propto N(2.82kT_{\rm s}).
     \label{eq:start_density}
\end{equation}

Assuming a fixed energy partition, the disc luminosity scales as $\dot{m}$, while the typical energy of a single seed photon scales as $\dot{m}^{1/4}$. Consequently, the seed photon number flux scales as:
\begin{equation}
     N(2.82kT_{\rm s})\sim \frac{L_d}{E_{\rm s}}\propto \dot m^{3/4}.
     \label{eq:seed_density}
\end{equation}
In the limit $E_{\rm s}\ll E_{\rm c}$, the photon number density near the turnover is
\begin{equation}
     N(E_{\rm s}) \approx N_0 E_{\rm s}^{-\Gamma}
     \label{eq:start_density2}
\end{equation}
Combining (\ref{eq:start_density}), (\ref{eq:seed_density}) and (\ref{eq:start_density2}), we obtain the scaling for the normalization factor:
\begin{equation}
     N_0 \propto \dot m^{3/4} E_{\rm s}^{\Gamma} \propto \dot{m}^{(3+\Gamma)/4}.
     \label{eq:normalization}
\end{equation}

The total upward-escaping Comptonized luminosity is given by:
\begin{equation}
     L_{\rm comp}^{\uparrow}\propto \int_{E_{\rm s}}^{\infty}E\,N(E)\,dE = N_0 \int_{E_{\rm s}}^{\infty} E^{1-\Gamma} \exp(-E/E_{\rm c}) dE,
     \label{eq:luminosity}
\end{equation}
in which the integral part can be expressed analytically as:
\begin{equation}
     \int_{E_{\rm s}}^{\infty} E^{1-\Gamma} \exp(-E/E_{\rm c}) dE = E_{\rm c}^{2-\Gamma} \Gamma_{\rm inc}(2-\Gamma, \xi),
\end{equation}
where $\xi = {E_{\rm s}}/{E_{\rm c}}$ and $\Gamma_{\rm inc}(s, \xi)$ denotes the incomplete gamma function. Utilizing this term, (\ref{eq:luminosity}) becomes:
\begin{equation}
     L_{\rm comp}^{\uparrow}\propto \dot{m}^{(3+\Gamma)/4} E_{\rm c}^{2-\Gamma} \Gamma_{\rm inc}(2-\Gamma, \xi).
     \label{eq:luminosity_final}
\end{equation}

With fixed effective disc-corona parameters, the upward Comptonized luminosity scales linearly with the accretion rate, i.e. $L_{\rm comp}^{\uparrow}\propto \dot m$. Equating this with (\ref{eq:luminosity_final}) yields an implicit expression for the $\Gamma-\dot{m}$ relation:
\begin{equation}
     E_{\rm c}^{2-\Gamma} \Gamma_{\rm inc}(2-\Gamma, \xi) \propto \dot{m}^{(1-\Gamma)/4}.
\end{equation}
Since $E_{\rm s}\propto \dot{m}^{1/4}$, we also have $\xi E_{\rm c}\propto \dot{m}^{1/4}$, so that:
\begin{equation}
     E_{\rm c}^{2-\Gamma} \Gamma_{\rm inc}(2-\Gamma, \xi) \propto \xi^{1-\Gamma} E_{\rm c}^{1-\Gamma}.
\end{equation}
Assuming no significant systematic correlation between $E_{\rm c}$ and $\dot{m}$, $E_{\rm c}$ can be treated as approximately constant. The implicit relation then reduces to
\begin{equation}
     \xi^{\Gamma-1}\Gamma_{\rm inc}(2-\Gamma, \xi) = {\rm const}.
     \label{eq:implicit}
\end{equation}

Since the high-energy cutoff is typically much larger than the seed-photon turnover energy, we have $\xi \ll 1$. For hard spectra with $1<\Gamma<2$, the incomplete gamma function $\Gamma_{\rm inc}(2-\Gamma, \xi)$ asymptotically approaches the Euler Gamma function $\Gamma_{\rm E}(2-\Gamma)$ (provided $(2-\Gamma)\,\Gamma_{\rm E}(2-\Gamma)\gg \xi^{2-\Gamma}$; otherwise, the analysis at $\Gamma\approx 2$ applies). Thus, (\ref{eq:implicit}) becomes:
\begin{equation}
     \xi^{\Gamma - 1}\Gamma_{\rm E}(2-\Gamma) = {\rm const}.
\end{equation}
Taking the logarithm and differentiating with respect to $\ln{\xi}$ gives:
\begin{equation}
     \frac{d\Gamma}{d\ln\xi} = \frac{\Gamma-1}{\psi(2-\Gamma)-\ln{\xi}}.
\end{equation}
where $\psi$ is the Digamma function. Since $\xi\ll 1$, $\ln{\xi}$ is a large negative number. For $\Gamma$ not too close to 2, the denominator is positive. Consequently, for hard spectra, $d\Gamma / d\ln\xi>0$. Because $\xi\propto \dot{m}^{1/4}$, this result proves a positive $\Gamma - \dot{m}$ correlation in the hard-spectrum regime.

For soft power-law spectra with $\Gamma>2$, $\Gamma_{\rm inc}(2-\Gamma, \xi)$ diverges and asymptotically scales as $\xi^{2-\Gamma}/(\Gamma-2)$. Equation (\ref{eq:implicit}) then reduces to:
\begin{equation}
     \frac{\xi}{\Gamma-2} = {\rm const},
\end{equation}
which indicates a linear relationship, $\Gamma - 2 \propto \dot{m}^{1/4}$, confirming a positive $\Gamma - \dot{m}$ correlation in the soft state.

Finally, at the critical transition where $\Gamma \approx 2$, the asymptotic limits above are no longer sufficient, so we differentiate the full implicit equation directly. Defining
\begin{equation}
     F(\Gamma, \xi) = \xi^{\Gamma - 1} \Gamma_{\rm inc}(2-\Gamma,\xi).
\end{equation}
The derivative $d\Gamma / d\ln \xi$ is given by:
\begin{equation}
     \frac{d\Gamma}{d\ln\xi} = -\frac{\xi\partial F / \partial \xi}{\partial F / \partial \Gamma}.
\end{equation}
The denominator is derived from the definition of the incomplete gamma function:
\begin{equation}
     F_{\Gamma} = \frac{\partial F}{\partial \Gamma} = -\xi \int_{1}^{\infty} (\ln{t}) t^{1-\Gamma} e^{-\xi t} dt.
\end{equation}
Since the integrand is strictly positive for $t> 1$, we have $F_\Gamma < 0$. And the numerator evaluated at $\Gamma=2$ is:
\begin{equation}
     F_{\xi} = \frac{\partial F}{\partial \xi}\Big|_{\Gamma=2} = E_1(\xi) - e^{-\xi},
\end{equation}
where $E_1(\xi)$ is the exponential integral function. Because $E_1(\xi) > e^{-\xi}$ strictly holds for $\xi < 0.4348$, we have $F_\xi > 0$. Therefore, $d\Gamma / d\ln\xi = - \xi F_{\xi} / F_\Gamma > 0$ holds near the critical point, ensuring that $\Gamma$ increases with $\dot{m}$ continuously across all physically relevant regimes.

\subsection{The correlation between \texorpdfstring{$\kappa_{2-10{\rm keV}}$}{kappa} and \texorpdfstring{$\lambda_{\rm Edd}$}{mdot}}

Compared to the $\Gamma-\dot{m}$ relation, the scaling of $\kappa_{2-10{\rm keV}}$ with $\dot{m}$ is more direct. Neglecting the subdominant Compton reflection component for simplicity, the 2-10 keV bolometric correction can be written as
\begin{equation}
     \kappa_{2-10{\rm keV}} = \frac{L_{\rm bol}}{L_{2-10}}=\frac{L_{\rm bol}}{L_{\rm comp}^{\uparrow}} \frac{L_{\rm comp}^{\uparrow}}{L_{2-10}}.
\end{equation}
Under the assumption that the effective coronal fraction and feedback geometry do not vary systematically with $\dot m$, the upward Comptonized luminosity scales approximately with the bolometric luminosity, so that
\begin{equation}
     \kappa_{2-10{\rm keV}} \propto \frac{L_{\rm comp}^{\uparrow}}{L_{2-10}}.
\end{equation}
Using the cutoff power-law approximation, this becomes
\begin{equation}
     \kappa_{2-10{\rm keV}} \propto \frac{\int_{E_{\rm s}}^{E_{\rm max}}E^{1-\Gamma}e^{-E/E_{\rm c}}dE}{\int_{2}^{10}E^{1-\Gamma}e^{-E/E_{\rm c}}dE},
\end{equation}
which integrates analytically to:
\begin{equation}
     \kappa_{2-10{\rm keV}} \propto \frac{\Gamma_{\rm inc}\left(2-\Gamma,\frac{E_{\rm s}}{E_{\rm c}}\right)-\Gamma_{\rm inc}\left(2-\Gamma,\frac{E_{\rm max}}{E_{\rm c}}\right)}{
\Gamma_{\rm inc}\left(2-\Gamma,\frac{2}{E_{\rm c}}\right)-\Gamma_{\rm inc}\left(2-\Gamma,\frac{10}{E_{\rm c}}\right)}.
\label{eq:kappa}
\end{equation}
As $\Gamma$ increases and the spectrum softens, a larger fraction of the radiative power is shifted below 2 keV, so the relative contribution of the fixed 2-10 keV band to the total coronal emission generally decreases. As a result, for the parameter range of interest here, $\kappa_{2-10{\rm keV}}$ is always expected to increase with $\Gamma$ in most cases. Since we have already shown that $\Gamma$ increases with $\dot{m}$, $\kappa_{2-10{\rm keV}}$ is likewise expected to show a positive trend with $\dot{m}$ over the relevant regime.

\begin{figure}
\includegraphics{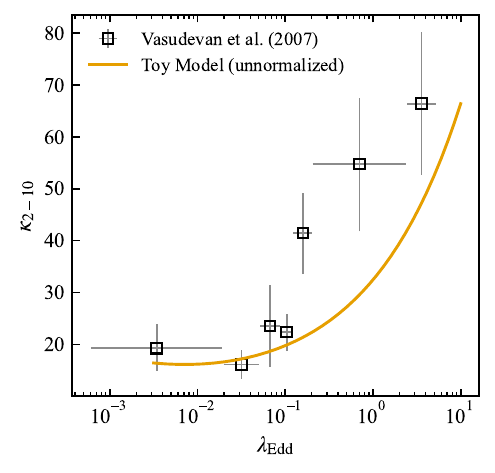}
\caption{Observed $\kappa_{2-10{\rm keV}}-\lambda_{\rm Edd}$ trend from \citet{2007MNRAS.381.1235V} compared with the toy-model prediction from (\ref{eq:kappa}). The model adopts $E_{\rm c} = 30$ keV, $E_{\rm s} = 0.1$, $E_{\rm max} = 500$ keV and the empirical $\Gamma-\dot{m}$ relation from \citet{2013MNRAS.433.2485B}. The toy-model curve is unnormalized and is shown only to illustrate that the positive trend can emerge even for fixed disc-corona energy partition fraction $f$ and the feedback factor $g$.}
\label{fig:toymodel}
\end{figure}

To illustrate this behavior, we adopt representative values of $E_{\rm c} = 30$ keV, $E_{\rm s} = 0.1$ keV, and an upper integration limit of $E_{\rm max} = 500$ keV. As an input $\Gamma-\dot{m}$ relation, we use the empirical fit reported by \citet{2013MNRAS.433.2485B},
\begin{equation}
     \Gamma = 2.34 + 0.34 \log_{10} \dot{m}
\end{equation}
Substituting this into (\ref{eq:kappa}), we obtain the predicted unnormalized $\kappa_{2-10{\rm keV}}-\lambda_{\rm Edd}$ trend shown in Figure \ref{fig:toymodel}. The resulting toy-model curve roughly reproduces the overall positive trend reported by \citet{2007MNRAS.381.1235V}, indicating that this behavior can arise even when the effective disc-corona energy partition and feedback geometry are fixed.

\section{Pair-balance calculation and constraints in slab coronae}
\label{sec:appendixE}

In classical slab-corona models, pair balance is often invoked as a thermostat-like mechanism that regulates the coronal temperature \citep[e.g.,][also see PS96]{1995ApJ...449L..13S}. Although the main conclusions of this paper do not rely on a pure-pair closure, we provide here a supplementary pair-balance calculation to connect our results with this classical interpretation and to assess whether a homogeneous purely thermal pair slab can plausibly supply the observed optical depths along the AGN locus.

In a corona dominated by electron-positron pairs, the positron and electron number densities are nearly equal and can be written as
\begin{equation}
     n_+ = n_- = \frac{\tau_{\rm T}}{2\sigma_{\rm T} H},
\end{equation}
where $H$ is the vertical scale height of the slab. The pair annihilation rate per unit volume is then given by \citep[e.g.,][]{1982ApJ...258..321S}:
\begin{equation}
\dot{n}_{\rm ann} = cr_{\rm e}^2 n_+ n_- \frac{\pi}{1+2\theta^2/\ln(1.3+2\eta_{\rm E}\theta)},
\end{equation}
where $r_{\rm e}$ is the classical electron radius and $\eta_{\rm E} = \exp(-\gamma_{\rm E}) \approx 0.5615$, with $\gamma_{\rm E}$ being the Euler-Mascheroni constant.

Electron-positron pairs are produced through photon-photon collisions between photons of energies $\epsilon_1$ and $\epsilon_2$. The threshold condition is
\begin{equation}
s= \frac{1}{2}\epsilon_1\epsilon_2(1-\cos\psi)>1,
\end{equation}
where $\psi$ is the angle between the photon propagation directions. In slab geometry,
\begin{equation}
\cos\psi=\mu_1\mu_2+\sqrt{(1-\mu_1^2)(1-\mu_2^2)}\cos\phi,
\end{equation}
where $\mu_1$ and $\mu_2$ are the direction cosines of two photons relative to the slab normal and $\phi$ is their relative azimuthal angle. Introducing the auxiliary variable
\begin{equation}
\zeta=\left(1-\frac{1}{s}\right)^{1/2},
\end{equation}
the corresponding Breit-Wheeler cross-section is
\begin{equation}
\sigma_{\gamma\gamma}
=\frac{3}{16}\sigma_{\rm T}(1-\zeta^2)
\left[
(3-\zeta^4)\ln\left(\frac{1+\zeta}{1-\zeta}\right)
-2\zeta(2-\zeta^2)
\right].
\end{equation}

For a given internal radiation field, we denote by
$I(\tau_{\rm T},\epsilon,\Omega)$ the specific intensity.
The corresponding differential photon number density is
\begin{equation}
     n_{\gamma} (\tau_{\rm T}, \epsilon, \Omega) = \frac{I(\tau_{\rm T}, \epsilon, \Omega)}{\epsilon m_{\rm e} c^3}.
\end{equation}
The local pair-production rate is computed via a multi-dimensional integral over photon energies and directions:
\begin{equation}
\begin{aligned}
    \dot{n}_{\gamma\gamma}=&\frac{c}{2}
\int d\epsilon_1\, d\epsilon_2\, d\Omega_1\, d\Omega_2\, \\
&\times\left\{
n_{\gamma}(\tau_{\rm T},\epsilon_1,\Omega_1)\,
n_{\gamma}(\tau_{\rm T},\epsilon_2,\Omega_2)\,
(1-\cos\psi)\,
\sigma_{\gamma\gamma}(s)\right\}.
\end{aligned}
\end{equation}
Pair equilibrium is reached when
\begin{equation}
\langle\dot{n}_{\gamma\gamma}\rangle=\dot{n}_{\rm ann},
\end{equation}
where the angle brackets denote an appropriate spatial average over height, i.e.
\begin{equation}
\langle\dot{n}_{\gamma\gamma}\rangle
=
\frac{1}{\tau_{\rm T}}
\int_0^{\tau_{\rm T}}
\dot{n}_{\gamma\gamma}(\tau)\, d\tau .
\end{equation}

To normalize the intensity of the radiation field, we define the compactness parameter corresponding to the effective coronal dissipated power, $L_{\rm diss}$,
\begin{equation}
l_{\rm diss}=
\frac{L_{\rm diss}\sigma_{\rm T}}{Hm_{\rm e}c^3}
\approx
2.3\times10^4
\left(\frac{L_{\rm diss}}{L_{\rm Edd}}\right)
\left(\frac{H}{R_{\rm g}}\right)^{-1}.
\end{equation}
For each chosen $(\theta, \tau_{\rm T})$, we use \texttt{compPSc} to compute the anisotropic internal radiation field $I(\tau_{\rm T}, \epsilon, \Omega)$ and then rescale its overall normalization using a trial $l_{\rm diss}$. Through substituting the rescaled field into the pair-production integral, we can iteratively solve for the critical compactness at which pair production balances annihilation.

\begin{figure*}
\includegraphics{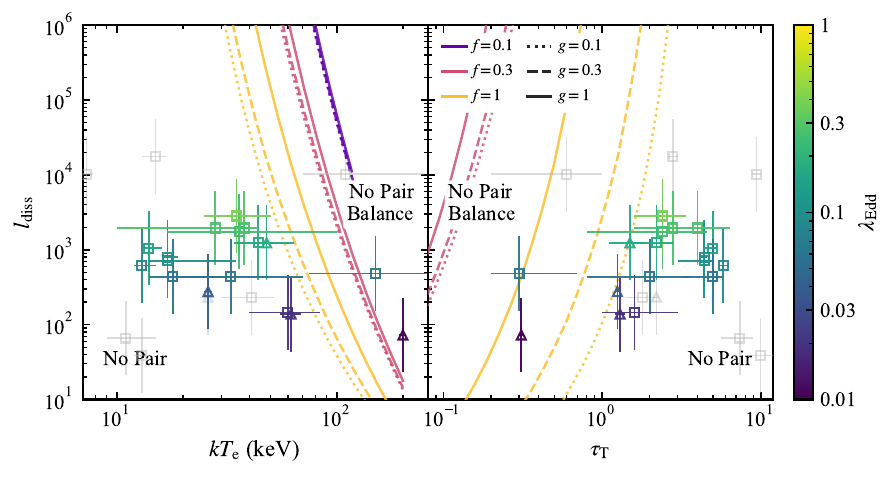}
\caption{Thermal pair-balance comparison for the observed slab-corona parameters. Left: critical compactness $l_{\rm diss}$ as a function of electron temperature $kT_{\rm e}$. Right: critical compactness $l_{\rm diss}$ as a function of optical depth $\tau_{\rm T}$. The observed sources are plotted using the same symbols and Eddington-ratio colour scale as in Figure \ref{fig:fitting}. Curves with different colours and line styles correspond to different coronal dissipation fractions $f$ and feedback parameters $g$. For each observed source, the vertical range of $l_{\rm diss}$ is estimated by assuming characteristic coronal sizes between $H=R_{\rm g}$ and $H=10R_{\rm g}$, with markers placed at the geometric mean.
\label{fig:pair}}
\end{figure*}

Previous works such as \citet{1995ApJ...449L..13S} focused mainly on the passive-disc limit, in which nearly all dissipated power is released in the corona and the cold disc only reprocesses the downward irradiation. Here we use the same pair-balance logic as a supplementary diagnostic, but allow the radiative-equilibrium solution to depend on the local coronal dissipation fraction $f$ and the feedback factor $g$. For fixed $(\theta, \tau_{\rm T})$, lowering $f$ increases the relative contribution of intrinsic disc photons, while increasing $g$ strengthens the photon feedback between disc and corona. Both effects cool the radiation field and therefore tend to increase the critical compactness required to maintain the same optical depth with pairs alone. The resulting critical pair-balance curves are shown in Figure \ref{fig:pair}. In both the $kT_{\rm e}-l_{\rm diss}$ and $\tau_{\rm T}-l_{\rm diss}$ planes, a pure thermal-pair equilibrium solution exists only along the corresponding critical curve. Below the curve, the radiation field is too weak to sustain the opacity with pairs alone, implying that an additional ion-electron component must contribute to the optical depth. By contrast, parameter combinations that yield compactness values substantially above the critical curve would tend to drive the system toward pair runaway in the pure-pair limit.

Since the main slab-equilibrium analysis favours $f\simeq1$, we plot the observational points using $L_{\rm diss}/L_{\rm Edd}\simeq \lambda_{\rm Edd}$ as an order-of-magnitude estimate. This choice is conservative for the pure-pair comparison, because taking $f<1$ would shift the observed points to lower $l_{\rm diss}$ and would make pair production even less effective at supplying the measured optical depths. We show the vertical range corresponding to characteristic slab heights $H=R_{\rm g}$ and $H=10R_{\rm g}$, with the plotted marker placed at the geometric mean of these two compactness estimates.

Figure \ref{fig:pair} shows that most of the low-temperature, moderately optically thick \texttt{compTT} sources lie below the pure-pair critical curves for representative reduced-feedback parameters. In particular, sources with $kT_{\rm e}\lesssim30\,{\rm keV}$ and $\tau_{\rm T}\gtrsim3$ would require critical compactness values exceeding $l_{\rm diss}\sim10^6$ in the pure-pair interpretation, far above the values inferred for plausible slab heights of $H\sim R_{\rm g}-10R_{\rm g}$. The same qualitative conclusion is visible in the $\tau_{\rm T}-l_{\rm diss}$ plane: except for the optically thin, high-temperature points, the observed optical depths are generally difficult to sustain with thermal pairs alone. We therefore conclude that, while pair balance remains a useful supplementary thermostat concept, the majority of the observed slab-corona locus is unlikely to correspond to a purely thermal, pair-dominated homogeneous slab. The fitted optical depths are more naturally interpreted as being dominated by an electron--ion plasma, with pairs possibly providing an additional contribution in some sources.

The supplementary \texttt{compPS} points provide a useful comparison but should not be over-interpreted. Except for the low-Eddington-ratio dim state of NGC~4151, which lies close to or beyond the pair-runaway side of the diagram, several \texttt{compPS} measurements fall closer to the pure-pair critical curves, especially in the $\tau_{\rm T}-l_{\rm diss}$ plane. This may indicate that individual objects such as NGC 4151, IC 4329A and MCG-5-23-16 are promising targets for more detailed source-specific studies of the pair content and geometry of the corona. It is worth noting that \textit{IXPE} observations of these sources have been interpreted as evidence for an extended disc-plane coronal geometry \citep{2023MNRAS.525.5437I,2023MNRAS.523.4468G,2024A&A...691A..29G, 2023MNRAS.525.4735T}. However, the compactness depends strongly on the assumed coronal size, and the inferred optical depth remains model-dependent. We therefore regard these \texttt{compPS} sources as useful future tests of the slab pair-balance picture, rather than as part of the primary evidence used to establish the constant-$y$ locus.


\bsp	
\label{lastpage}
\end{document}